\documentclass[aps,amsmath,amssymb, notitlepage, 
twocolumn,
nofootinbib,
]{revtex4-1}

\usepackage[pdftex]{graphicx}
\usepackage{dcolumn}
\usepackage{bm}
\usepackage{epsfig}
\usepackage{latexsym}
\usepackage{amsmath}
\usepackage{amsfonts}
\usepackage{amssymb}
\usepackage{color}
\usepackage{array}
\usepackage{framed}
\usepackage{esvect}
\usepackage{caption}
\usepackage{subcaption}
\usepackage{slashed}

\renewcommand{\v}[1]{\mathbf{#1}} % \v -> vector (bf)

\newcommand{\be}{\begin{equation}}
\newcommand{\ee}{\end{equation}}
\newcommand{\bea}{\begin{eqnarray}}
\newcommand{\eea}{\end{eqnarray}}

\renewcommand{\vec}[1]{{\bf #1}}
\renewcommand{\hat}[1]{{\widehat #1}}

\begin{document}

\title{ Half-filled Landau level, topological insulator surfaces, and three dimensional quantum spin liquids}
\author{Chong Wang  and T. Senthil}
\affiliation{Department of Physics, Massachusetts Institute of Technology,
Cambridge, MA 02139, USA}
\date{\today}
\begin{abstract}
We  synthesize and partly review recent developments relating the physics of the half-filled Landau  level in two dimensions to correlated surface states of topological insulators in three dimensions. The latter are in turn related to the physics of certain three dimensional quantum spin liquid states.  The resulting insights provide an interesting answer to the old question of how particle-hole symmetry is realized in composite fermion liquids. Specifically  the metallic state at filling $\nu = \frac{1}{2}$ - described originally in pioneering work by Halperin , Lee, and Read as a liquid of composite fermions - was proposed recently by Son to be described by a particle-hole symmetric effective field theory distinct from that in the prior literature. We show how the relation to topological insulator surface states leads to  a physical understanding of the correctness of this proposal. We develop a simple picture of the particle-hole symmetric composite fermion through a modification of older pictures as electrically neutral ``dipolar" particles.  We revisit the phenomenology of composite fermi liquids (with or without particle-hole symmetry), and show that their heat/electrical transport dramatically violates the conventional Wiedemann-Franz law but satisfies a modified one.  We also discuss the implications of these insights for finding physical realizations of correlated topological insulator surfaces. 

\end{abstract}
\maketitle

\tableofcontents

\section{Introduction}
Recently a number of seemingly disparate research topics have converged and have been seen to be closely related to each other. The first is the classic problem of a half-filled Landau level of spin-polarized electrons in two space dimensions\cite{dassarmabook,jainbook}. The second is the effects of interactions on three dimensional topological insulators, and in particular, the possibility of novel strongly correlated surface states  of such insulators\cite{tsarcmp}. The third is the study of three dimensional quantum spin liquid phases with an emergent gapless photon, such as may possibly be realized in quantum spin ice materials\cite{gingrasrev}.  As expected from such a convergence new insights on each of these problems have emerged. Amongst other results, an old issue in the theory of the half-filled Landau level now has a simple and elegant answer.   In a different direction, correlated surface states of some three dimensional topological insulators are now seen to have a surprising physical realization in  ordinary two dimensional systems. 

The purpose of this article is to synthesize and elaborate on  these developments.  The core of what we describe is based on several recent papers\cite{sonphcfl,tsymmu1,dualdrcwts2015,dualdrcmaxav}. However the point of view and emphasis that we provide is different from what is contained in these papers and other existing literature. We present a simplified, and physically transparent perspective, that distills the essence of the ideas involved.

We begin by describing the three different research topics separately. 

\subsection{The half-filled Landau level}
Electrons confined to two dimensions in a strong magnetic field display the phenomenon of the integer and fractional quantum Hall effects. We will be concerned with an ``unquantized" quantum Hall effect (see for eg, the contribution by Halperin in Ref. \onlinecite{dassarmabook}) that 
occurs when the filling factor $\nu$ of the lowest Landau level is $\frac{1}{2}$. Empirically this is seen to be a metal albeit a rather unusual one.  The classic theory of this metal - due to Halperin, Lee, and Read (HLR)\cite{hlr} - describes this as a compressible state   obtained by forming a fermi surface of ``composite fermions"\cite{jaincf} rather than the original electrons. In the original HLR theory , the composite fermions are formed by binding two flux quanta to the physical electrons. At $\nu = \frac{1}{2}$  this attached flux on average precisely cancels the external magnetic flux so that the  composite fermions move in effective zero field. This facilitates the formation of a Fermi surface and leads to an effective field theory of the metal as a Fermi surface coupled to a fluctuating gauge field which is then used to describe the physical properties of this metal. 

The HLR theory - and some subsequent refinements -  successfully predicted many experimental properties.  For instance when the filling is tuned slightly away from $\frac{1}{2}$, the composite fermions see a weak effective magnetic field and their trajectories are expected to follow cyclotron orbits with radii much larger than the underlying electrons.  These have been directly demonstrated in experiment\cite{willett93,kang93,goldman94,smet96} - for reviews see, e.g., the contribution by Tsui and Stormer in Ref. \onlinecite{dassarmabook}, and Ref. \onlinecite{willett97}. Further the composite Fermi liquid acts as a parent for the construction of the Jain sequence of states\cite{jainbook} away from $\nu = \frac{1}{2}$: they are simply obtained by filling an integer number of Landau levels of the composite fermions. Finally the composite Fermi liquid yields the non-abelian Moore-Read quantum Hall state  through pair ``condensation" of the composite fermions\cite{readgrn}. 

Despite its success there was one unresolved question with the theory of the composite Fermi liquid at $\nu = \frac{1}{2}$. To appreciate this consider the limit that  the Landau level  spacing $\hbar \omega_c  \gg H_{int}$ (where $\omega_c$ is the cyclotron frequency and $H_{int}$ is the electron-electron interaction). Then it is legitimate to project to the lowest Landau level. With a two-body interaction (e.g., Coulomb) the resulting Hamiltonian has a ``particle-hole" symmetry at $\nu = \frac{1}{2}$. This symmetry is not manifest in the HLR description of the composite fermi liquid , and is possibly even violated by it\cite{klkgphhlr,maissamph}. A lowest Landau level description is often routinely used in theoretical discussions and numerical calculations of quantum hall states, including at $\nu = \frac{1}{2}$.  It is also not an unrealistic limit to consider in experiments. It is thus important to understand how the particle-hole symmetry should be incorporated into the theory of the composite Fermi liquid. 

\subsection{Interacting topological insulators in three dimensions}
In the last decade condensed matter physics has been invigorated by the study of topological insulating phases of matter\cite{HsnKn,HsnMr,QiScz}. While much of the initial theoretical discussion focused on models of non-interacting electrons, in recent years attention has turned to  studies of the phenomenon of topological insulation in strongly interacting electronic systems.   The effects of interactions raises many questions.  Is the topological distinction between phases obtained in free fermion models robust to the inclusion of interactions? Are there new phases enabled by interactions that have no free fermion description? Even if a free fermion topological phase survives in an interacting system, are there new correlated surface states that can appear as an alternate to the ones obtained in the free fermion model? 

Tremendous progress on these questions has been achieved theoretically. Our focus here is on three dimensional topological insulators (TI). In that case for spin-orbit coupled insulators the free fermion topological insulator is known to be stable to interactions\cite{qitheta}. Within band theory the surface of such an insulator famously consists of the odd number of electronic Dirac cones. This metallic surface cannot be gapped or rendered insulating with any amount of impurities so long as the defining symmetries (charge conservation and time reversal) are preserved. On the other hand with interactions several groups\cite{fSTO1,fSTO2,fSTO3,fSTO4} described how a symmetry preserving gapped surface can emerge for the bulk topological insulator. Inspired by similar constructions\cite{avts12,hmodl,statwitt,burnell,geraedts14} for bosonic analogs of the topological insulators these papers showed that such a symmetry preserving gapped surface requires the kind of topological order familiar from discussions of the fractional quantum Hall effect and some quantum spin liquids.  However the symmetry is implemented in this topologically ordered state in a manner that is forbidden (``anomalous") in a strictly two dimensional system. Symmetry-preserving surface topologically ordered phases, besides being conceptually interesting, proved to be a useful theoretical tool in describing the physics of a class of interacting generalizations of topological insulators known as Symmetry Protected Topological (SPT) phases\cite{chencoho2011}.  These are phases with no non-trivial bulk excitations but which nevertheless have  non-trivial surface states protected by a global symmetry.  

Spin-orbit coupled electronic SPT insulators in $3d$ have a classification\cite{3dfSPT} by the group $Z_2^3$ as compared to the $Z_2$ classification without interactions. In interacting systems there are thus  6 spin-orbit coupled SPT insulators in 3d that are `beyond band theory'. Electronic SPT phases with many physically interesting symmetry groups in $3d$ have been classified\cite{fidkowski3d,3dfSPT,3dfSPT2} and their properties are understood. In several symmetry classes there exist SPT phases which are `beyond band theory' ({\em i.e} have no free fermion description)\cite{3dfSPT,3dfSPT2}.  In addition  for some symmetries some free fermion topological phases become indistinct from topologically trivial phases in an interacting system\cite{fidkowski3d,3dfSPT,3dfSPT2,maxvortex}.  Thus the classification of $3d$ free fermion SPT phases is modified in the presence of interactions (see Ref.\onlinecite{tsarcmp} for a review). 

An important open question in this area is the physical realization of these various phenomena.  For instance what kinds of physical systems naturally realize the correlated surface states of the three dimensional topological insulator?   

 \subsection{Quantum spin liquids in three dimension}
Quantum spin liquids are  ground states of  interacting quantum spin systems characterized by long range entanglement between local degrees of freedom. While the theoretical possibility of such ground states has been appreciated for a long time it is only in the last decade that credible experimental candidates have emerged\cite{palrev}. There are many  kinds of quantum spin liquid phases which are sharply distinct from each other. Of particular interest to us are three dimensional quantum spin liquid phases that possess an emergent gapless photon in the excitation spectrum\cite{wenbook,bosfrc3d,wen03,3ddmr,hfb04,lesikts05,kdybk,shannon}. The low energy theory of such phases is a (deconfined) $U(1)$ gauge theory. These phases  are hence called $U(1)$ quantum spin liquids. Their excitation spectrum consists of a gapless emergent `photon', and emergent particle-like excitations that couple to the photon as electric or magnetic charges. Such spin liquids may possibly be realized in quantum spin ice materials on pyrochlore lattices\cite{balentsqspice}. The spin hamiltonian describing these pyrochlore magnets is rather complicated and is characterized by very little symmetry\cite{balentsqspice}. The only internal symmetry is time reversal. This motivates a classification and description of time reversal invariant $U(1)$ quantum spin liquids in three dimensions\cite{tsymmu1}. 

Of particular interest to us is the so-called ``Topological Mott Insulator" discussed in Ref. \onlinecite{pesinlb} as a possible state in pyrochlore iridates such as $Y_2Ir_2O_7$. This is a three dimensional time reversal symmetric $U(1)$ quantum spin liquid state where the gapped emergent `electric'  charge (denoted a spinon) is a gapped fermion that is a Kramers doublet under the time reversal symmetry. Furthermore this spinon has topological band structure leading to protected surface states. 

Naively many other similar constructions of $U(1)$ quantum spin liquids are possible where the emergent electric or magnetic charges themselves form an SPT phase. How are these different constructions related to each other? 

\subsection{Summary and plan}
We will see below that these three topics are closely connected to each other, and that these connections lead to a wealth of fresh insights.  In a recent paper Son\cite{sonphcfl} proposed a particle-hole symmetric formulation of the composite Fermi liquid in the half-filled Landau level. This proposal was motivated by thinking about the half--filled Landau level in a microscopic system of Dirac fermions in a magnetic field such as may arise at the surface of a three dimensional topological insulator.   The composite fermi liquid was suggested to be also described by a single Dirac cone at finite density (with particle-hole symmetry playing the role of time reversal), and with a coupling to a $U(1)$ gauge field.  In another recent paper\cite{tsymmu1} the present authors classified and described the physics of time reversal symmetric $U(1)$ quantum spin liquids in $3d$. The results were then applied in Ref. \onlinecite{dualdrcwts2015} to deriving a new gapless metallic surface state of the $3d$ topological insulator. This same result and some of the results of Ref. \onlinecite{tsymmu1} were also independently obtained in Ref. \onlinecite{dualdrcmaxav}. The improved understanding of the topological insulator surface paves the way for an understanding of Son's proposal. 

In the rest of this paper we will synthesize these results in a manner that exposes the physics most simply.  We begin by describing the action of particle-hole symmetry in the half-filled Landau level (Sec. \ref{phhll}) and then describe  Son's proposed theory (Sec. \ref{sonprop}). Next, in Sec. \ref{phcflphys} we provide a physical description of the particle-hole symmetric composite fermion.  This modifies and extends a previous physical picture of the composite fermion as a neutral dipolar particle. We argue that this modification is natural when particle-hole symmetry is present. We show that the most essential features of the particle-hole symmetric composite fermion follow simply and naturally from this modified picture. We support our arguments by solving in Appendix \ref{drcdip} a simple model of two-particle quantum mechanics which illustrates several of the key features. We then provide, in Secs. \ref{hlltisrf} and \ref{cfsti}, an alternate understanding of the half-filled Landau level by relating it to correlated surface states of three dimensional fermionic topological insulators.  As described above such surface states have ``anomalous" symmetry implementation not possible in a strictly $2d$ system. Remarkably 
the well studied half-filled Landau level -despite being strictly $2d$ - provides a physical realization of such a state, and makes it relevant to experiments. This is possible because the particle-hole symmetry is not really a microscopic local symmetry in the physical Hilbert space of the two dimensional system but is an emergent low energy symmetry of a single Landau level. 

We describe (in Sec. \ref{titsymmu1}) how these correlated surface states are fruitfully constrained by studying the properties of the three dimensional bulk when the fermions are coupled to a dynamical $U(1)$ gauge field. The resulting state is to be viewed as a $3d$ $U(1)$ quantum spin liquid - in particular a ``topological Mott insulator". We review arguments of Refs. \onlinecite{tsymmu1,dualdrcmaxav} showing that the topological Mott insulator admits two equivalent but dual descriptions as either charge or monopole topological insulators in Sec. \ref{bdgti}. The consequences\cite{dualdrcwts2015,dualdrcmaxav} of this bulk duality for correlated surface states of the original topological insulator are then described in Secs. \ref{dualsrf}  and \ref{vms}. We then revisit the composite fermi liquid (in Sec. \ref{bcfl}) with this understanding of the
correlated surface states and show that it matches exactly with Son's proposed theory.  We then consider (Sec. \ref{phpf}) a particle-hole symmetric version of a paired non-abelian quantum Hall state\cite{sonphcfl} obtained by pairing the composite fermions. This state is identical to a symmetry preserving surface topologically ordered state discussed previously\cite{fidkowski3d,3dfSPT2,maxvortex} for the corresponding $3d$ topological insulator.  We show that this particle-hole symmetric Pfaffian state gives further support to the modified dipolar picture of the composite fermion. 

With this understanding we revisit the phenomenology of composite Fermi liquids in Sec. \ref{cflphen} with or without particle-hole symmetry. We show that many of the essential features of the HLR theory (which have successfully confronted experiment) are preserved, for instance in the electromagnetic response. We turn next to heat transport of  the composite fermi liquid metal (which does not seem to have been discussed before).  We show, both within the conventional HLR theory and the new particle-hole symmetric version,  that there is a dramatic violation of the conventional Wiedemann-Franz relationship between the heat and electrical conductivities. However the composite fermi liquid should satisfy a modified Wiedemann-Franz law. This can possibly be tested in future experiments. We also make some brief comments on the cyclotron radius away from half-filling, and on the effects of disorder. A key feature of the particle-hole symmetric theory is the presence of a $\pi$ Berry phase when the composite fermion circles around the Fermi surface. In Appendix \ref{sdhpi} we show that this Berry phase is implied by the standard Shubnikov-deHaas oscillations near $\nu = \frac{1}{2}$ after a simple but revealing reinterpretation.

 \section{Particle-hole symmetry and the half-filled Landau level}
 \label{phhll}
 We begin with the half-filled Landau level in two dimensions and describe the action of particle-hole symmetry. 
 Consider the full set of single particle eigenstates $\phi_{I,m} (x,y)$ where $I$ labels the Landau level and $m$ the orbital within each Landau level, for instance, in the symmetric gauge. The microscopic electron destruction operator $\psi_e(x,y)$ 
 may be expanded as
 \begin{equation}
 \psi_e(x,y) = \sum_{I,m} \phi_{I,m}(x,y) c_{I,m}
 \end{equation}
 The $c_{I,m}$ are electron destruction operators for the single particle state indexed by $(I,m)$ and satisfy the usual fermion anti commutation relations. To project to the lowest Landau level we truncate the expansion by keeping only the $I =0$ terms: 
 \begin{equation}
 \psi_e(x,y) \approx \sum_m \phi_{0m} c_m
 \end{equation}
 (Here and henceforth drop the Landau level index $0$ and denote $c_{0,m}$ simply by $c_m$).  The particle-hole transformation in the lowest Landau level is defined to be an {\em anti-unitary} operator $C$ such that 
 \begin{eqnarray}
 Cc_m C^{-1} & = & h_m^\dagger \\
 Cc_m^\dagger C^{-1} & = & h_m
 \end{eqnarray}
 The $h_m$ satisfy fermion anti commutation relations.  A two-body Hamiltonian acting within the lowest Landau level can be written
 \begin{equation}
 H_{int} = \frac{1}{2} \sum_{m_1, m_2, m_1', m_2'} c^\dagger_{m_1'} c^\dagger_{m_2'} c_{m_2} c_{m_1} \langle m_1' m_2' | V m_1 m_2 \rangle
 \end{equation}
 The anti-unitary  $C$ operation leaves this interaction invariant but generates a one-body term. At half-filling this is exactly compensated by a chemical potential so that the Hamiltonian is particle-hole symmetric. 
 Note that the total electron number $N_e = \sum_m c_m^\dagger c_m$ transforms as 
 \begin{equation}
 C(\sum_m c_m^\dagger c_m ) C^{-1} = N_\phi - \sum_m h_m^\dagger h_m 
 \end{equation}
 ($N_\phi$ is the number of flux quanta and hence the degeneracy of the Landau level). Thus as expected the electron filling factor $\nu = \frac{N_e}{N_\phi}$ transforms to $1 - \nu_h$ with $\nu_h$ the hole filling factor. 
 
 Note that under the $C$ transformation the empty state $|0 \rangle$ is transformed to the filled Landau level. If a state $|\Psi \rangle$ at $\nu = \frac{1}{2}$ is particle-hole invariant, {\em i.e} $C\Psi \rangle = |\Psi\rangle$, 
 then we can view it either as a state of electrons at half-filling or as the combination of a filled Landau level and the same state of holes at half-filling. This leads to the conclusion\cite{klkgphhlr} that the electrical Hall conductivity in such a state is exactly $\sigma_{xy} = \frac{e^2}{2h}$. 
 
The full symmetry of the half-filled Landau level thus is $U(1) \times C$ (the $U(1)$ is the familiar charge-conservation symmetry).  As $C$ is anti-unitary, the direct product structure means that the generator of $U(1)$ rotations (the deviation of the physical charge density from half-filling) is odd under $C$. 

\section{Particle-hole symmetry and the  composite fermi liquid} 
\label{sonprop}
It has been appreciated for some time\cite{klkgphhlr,maissamph} that the effective field theory proposed by HLR for the half-filled Landau level is not manifestly particle-hole symmetric, and is perhaps even inconsistent with it. On the other hand numerical calculations performed in the lowest Landau level show that with the projected 2-body Coulomb interaction the Fermi-liquid like state at half-filling preserves particle-hole symmetry(see for instance \cite{rzyhald2000}). It is therefore important to construct a description of the composite fermi liquid theory which explicitly preserves the particle-hole symmetry. A very interesting proposal for such a theory was made recently by Son\cite{sonphcfl}. The composite fermion was proposed to be a two-component Dirac fermion field $\psi_v$ at a finite non-zero density, and with the effective (Minkowski) Lagrangian: 
\begin{equation}
\label{ccfl}
 {\cal L}  =   i\bar{\psi}_v \left(\slashed{\partial} + i \slashed{a}\right) \psi_v - \mu_v \bar{\psi}_v \gamma_0 \psi_v+ \frac{1}{4\pi} \epsilon_{\mu\nu\lambda} A_\mu\partial_\nu a_\lambda \\
 \end{equation}
 Here $a_\mu$ is a fluctuating internal $U(1)$ gauge field and $A_\mu$ is an external probe gauge field. The $2 \times 2$ $\gamma$ matrices are $\gamma_0 = \sigma_y, \gamma_1 = i\sigma_z, \gamma_2 = -i\sigma_x$.   $\mu_v$ is a composite fermion chemical potential that ensures that its density is non-zero. The physical electric current is 
 \begin{equation}
 \label{dualj}
 j_\mu =  \frac{1}{4\pi} \epsilon_{\mu\nu\lambda} \partial_\nu a_\lambda
 \end{equation}
 Here the $0$-component is actually the deviation of the full charge density $\rho$ from that appropriate for half-filling the Landau level, {\em i.e}
 \begin{equation}
 \label{jorho}
 j_0 = \rho - \frac{B}{4\pi}
 \end{equation}
 Here and henceforth (unless otherwise specified) we will work in units where the electron charge $e = 1$ and $\hbar = 1$.  In the presence of  long range Coulomb interactions, the above Lagrangian must be supplemented with an additional interaction term $ \int_{\vec x, \vec x'} j_0(\vec x) V(\vec x - \vec x') j_0 (\vec x')$ where $V$ is the Coulomb potential. 
 
 The Lagrangian above describes  the dynamics of the composite fermions, and their coupling to external probe electromagnetic fields. To obtain the full response 
 of the lowest Landau level to the electromagnetic field, this Lagrangian must be supplemented by a `background' Chern-Simons term which accounts for the $\sigma_{xy}
 = \frac{e^2}{2h}$ demanded by particle-hole symmetry. This background term takes the form
 \begin{equation}
 \label{lbg}
 {\cal L}_{bg} = \frac{1}{8\pi} \epsilon_{\mu\nu\lambda} A_\mu \partial_\nu A_\lambda
 \end{equation}
 
Note the similarity of Eqn. \ref{dualj} with the usual HLR theory.  There are however some important differences between Son's proposal and the HLR theory.  Under the original particle-hole symmetry operation $C$, the composite fermion field $\psi_v$ is hypothesized to transform as 
 \begin{equation}
 \label{Cpsiv}
 C\psi_v C^{-1} = i\sigma_y \psi_v
 \end{equation}
 Thus $\psi_v$ goes to itself rather than to its antiparticle under $C$. Further this transformation implies that the two components of $\psi_v$ form a Kramers doublet under $C$ (recall that $C$ is anti unitary).  
 With this transformation, the Lagrangian is manifestly invariant under $C$ so long as we choose $a_0 \rightarrow a_0, a_i \rightarrow - a_i$ and let the time $t \rightarrow - t$.  The deviation $j_0$ of the physical charge density from half-filling   (see Eqn. \ref{dualj}) is then odd under $C$ as required. 

These composite fermions are at a non-zero density $\frac{B}{4\pi}$ and fill states upto a Fermi momentum $K_f$.  This should be compared with the HLR theory where the prescription for the composite fermion density is just the electron density $\rho$. At half-filling we have $\rho = B/4\pi$ and the two  prescriptions agree. . However these two 
prescriptions are different on going away from half-filling. We will see later (in Appendix \ref{sdhpi}) that this slight difference actually plays a crucial role. 

Returning to the particle-hole symmetric theory, the ``Diracness" of the composite fermion is manifested  as follows: when a composite fermion at the Fermi surface completes a full circle in momentum space its wave function acquires a Berry phase of $\pi$.  This is a ``low-energy" manifestation of the Dirac structure that does not rely on the specifics of the dispersion far away from the Fermi surface.

Finally notice that unlike in the original HLR theory (but actually similar to subsequent work\cite{read98,avl1} on the related problem of bosons at $\nu = 1$) there is no Chern-Simons term for the internal gauge field $a_\mu$. 

If we ignore the gauge field, Eqn. \ref{ccfl} actually describes a single Dirac cone that arises at the surface of $3d$ spin-orbit coupled topological insulators. Interestingly, in this effective theory,  $C$ plays the role of time reversal as is clear from Eqn. \ref{Cpsiv}.  Thus the proposed particle-hole symmetric composite fermi liquid theory is this single Dirac cone coupled to an emergent  $U(1)$ gauge field. 

In the sections that follow we will build an understanding of the correctness of  Son's proposal through physical arguments and by relating the half-filled Landau level to topological insulator surface states.  An alternate recent discussion\cite{maissamph} of particle-hole symmetry in the half-filled Landau level proposes an ``anti-HLR" state as a particle-hole conjugate of the HLR state.  We will however not describe it here. 

 \section{Physical picture of the particle-hole symmetric composite fermion}
 \label{phcflphys}
 We now provide a very simple physical picture of these particle-hole symmetric composite fermions by relating them to previous constructions of the composite Fermi liquids. Subsequent to the original HLR theory through a process of intense reexamination\cite{read94,rsgm97,Pasquier1998,read98,dhleephcf98,sternetal99,gmrsrmp03} a picture of the composite fermion as a neutral dipolar particle emerged. This is illustrated by considering the composite fermion at a filling $\nu = \frac{p}{2p+1}$ slightly different from $\frac{1}{2}$.  Then a fractional quantum hall state is possible and is described by filling $p$ Landau levels of microscopic composite fermions obtained by the usual attachment of $4\pi$ flux to the electron. At the mean field level the excitations about this state are single microscopic composite fermions but their charge/statistics will be modified by the background quantum hall effect.  The true low energy quasiparticle has fractional charge 
 $e^* = \frac{e}{2p+1}$. Thus when $\nu$ goes to $\frac{1}{2}$ (corresponding to $p$ going to $\infty$), the low energy quasiparticle might be expected to have $e^* = 0$. Its statistics also reverts back to fermionic when $p \rightarrow \infty$. 
 
\begin{figure}
\begin{center}
\includegraphics[width=1.9in]{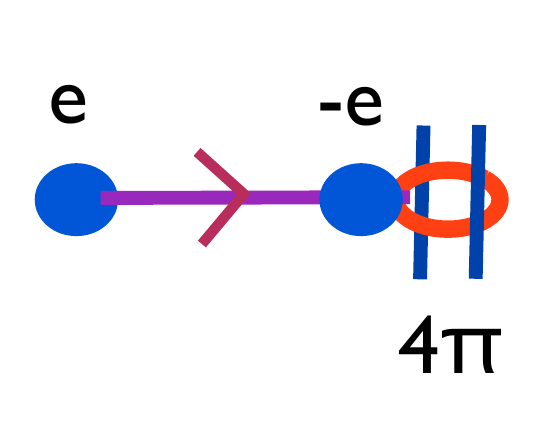}
\end{center}

\caption{ The standard picture of the composite fermion at $\nu = \frac{1}{2}$ regards it as an electron (of charge $e$) bound to a $4\pi$ vortex. The vortex carries charge $-e$ and is displaced from the electron in the direction perpendicular to its momentum. The composite fermion is thus viewed as a neutral dipolar particle. }
\label{olddip}

\end{figure}

 Physically due to the electrical Hall conductivity $1/2$, the $4\pi$ flux attached to the electron acquires an electric charge of $-e$ which compensates for the electron's charge. 
 In a lowest Landau level description of the theory, it is appropriate to replace the concept of flux attachment with the related concept of binding vortices to the particles. In such a description Read proposed\cite{read94}, based on a wave function for the HLR state, that the vortex is displaced from the electron by an amount perpendicular to the momentum of the composite fermion.  The key idea is that when projected to the lowest Landau label a phase factor like $e^{i\vec k \cdot \vec r}$ generates a translation of the correlation hole (the vortex) bound to the electron by an amount proportional to and perpendicular to the momentum.  Let us briefly describe this logic. 
 The standard flux attachment procedure leads naturally to a wave function  for the composite Fermi liquid: 
\begin{equation}
\psi(z_1, ........, z_N) = P_{LLL} det(e^{i\vec k_i \cdot \vec r_j}) \prod_{i < j} (z_i - z_j)^2  
\end{equation}
Here $z_i$ are the complex coordinates of the ith electron and $\vec r_i$ is the same coordinate in vector form. We have suppressed the usual Gaussian  factors. 
This is known as the Rezayi-Read wave function\cite{rzyrd94}. The factor $(z_i - z_j)^2$ has the effect of attaching a $4\pi$ vortex to each electron to convert it into a composite fermion. The Slater determinant then builds a Fermi sea of the composite fermions. As the plane wave factors do not stay within the lowest Landau level it is necessary to project back to it through the operator $P_{LLL}$.  Now write
\begin{equation}
e^{i\vec k \cdot \vec r} = e^{\frac{i}{2} \left( \bar{k} z + k \bar{z} \right)}
\end{equation} 
Here $k = k_x + ik_y$ and $\bar{k} = k_x - ik_y$. 
In the lowest Landau level we should replace ${\bar z} \rightarrow 2l_B^2 \frac{\partial}{\partial z}$. This leads to the expectation that in the wave function the terms involving $\bar{z}$ will shift the vortex away from the particle in the direction perpendicular to $\vec k$ and by an amount proportional to it.  This line of thought leads to a dipolar picture of the composite fermion as shown in Fig. \ref{olddip}.   
 
Though this wave function -based thinking has been criticized (see e.g. Ref. \onlinecite{gmrsrmp03}), the final dipolar description gained wide acceptance in the late 90s through more sophisticated  kinds of calculations\cite{rsgm97,Pasquier1998,read98,dhleephcf98,sternetal99}. 
 
The resulting picture was that the low energy composite fermions (as opposed to the microscopic composite fermions) were electrically neutral dipoles with a dipole moment perpendicular to their momentum (see Fig. \ref{olddip}), and it is these low energy composite fermions that live near the Fermi surface. These neutral dipolar composite fermions continue to couple to a $U(1)$ gauge field but without a Chern-Simons term. The flux of this gauge field however is the physical electrical 3-current and hence couples directly to the external probe gauge field.     

Particle-hole symmetry was not addressed in these prior works (except by Dung-Hai Lee's work\cite{dhleephcf98} whose exact relation with the present circle of ideas is not clear).  Here we show how a modification of this picture captures the essential features of the particle-hole symmetric composite fermion.

 \begin{figure}
\begin{center}
\includegraphics[width=1.9in]{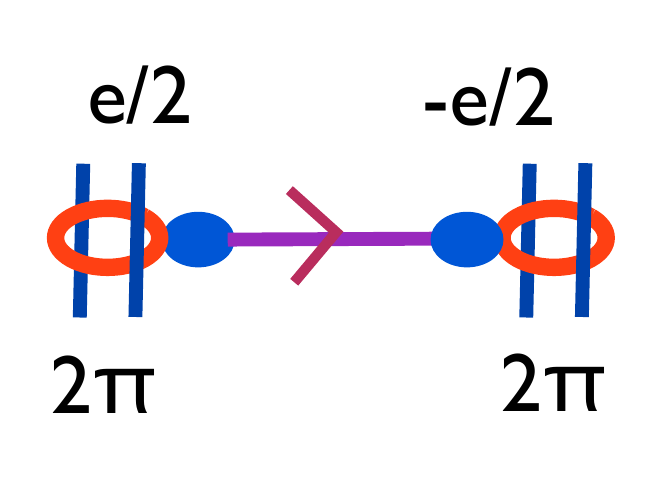}
\end{center}

\caption{ The new picture of the particle-hole symmetric composite fermion at $\nu = \frac{1}{2}$.  One end of the dipole has a $2\pi$ vortex bound to charge $\frac{e}{2}$. The other end has a charge $- \frac{e}{2}$ also bound to a $2\pi$ vortex.  The  displacement between the two is in the direction perpendicular to their center-of-mass momentum.  The positively charged end can be viewed as a $2\pi$ vortex located exactly on the electron. Thus compared to the picture in Fig. \ref{olddip} only one $2\pi$ vortex is displaced from the electron. }
\label{newdip}

\end{figure}

 \begin{figure}
\begin{center}
\includegraphics[width=1.9in]{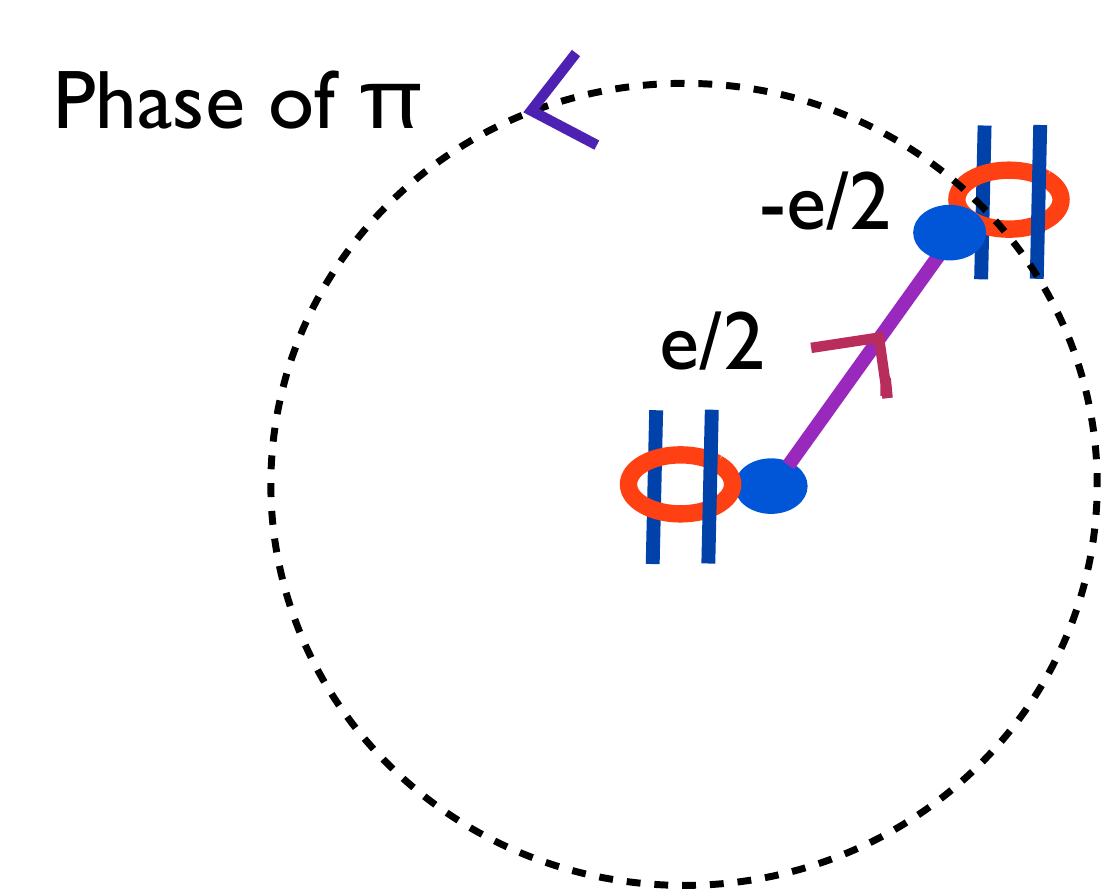}
\end{center}

\caption{When one end of the dipole of Fig. \ref{newdip} is rotated in a closed loop around the other end, there is a phase of $\pi$.}
\label{dippiphs}

\end{figure}

Let us begin with a discussion of wave functions for the half-filled Landau level which was the initial motivation for the dipolar picture.  
We now show how this line of thinking leads actually to a different picture which naturally enables a particle-hole symmetric description, and provides a physical basis to Son's proposal. 

It is well known that fermion wave functions in the lowest Landau level must have the structure
\begin{equation}
\psi(z_1, z_2, ......, z_N) = \prod_{i < j} (z_i - z_j)f(z_1, ......., z_N)
\end{equation}
where $f$ is a symmetric polynomial. The $z_i - z_j$ structure is a zero of the wave function that is demanded by Pauli exclusion.  Thus whatever state we build in the lowest Landau level, Pauli exclusion guarantees that there is one $2\pi$ vortex that is sitting exactly on top of the electron.

At $\nu = \frac{1}{2}$, the symmetric function $f$ can be taken to be the wave function of bosons at $\nu = 1$ which can also form a  composite Fermi liquid state. For bosons at $\nu = 1$ the composite Fermi liquid theory is, in fact, better established theoretically\cite{Pasquier1998,read98,avl1} than for fermions at $\nu = \frac{1}{2}$. This bosonic composite liquid is obtained by binding a $2\pi$ vortex to the particle. Wavefunction, or other arguments, then show that this vortex is indeed displaced from the particle in the  manner described above. 

We thus expect the following picture for the structure of the composite fermion at $\nu = \frac{1}{2}$. One $2\pi$ vortex sits exactly on the electron while the other is displaced from it (in the direction perpendicular to the composite fermion momentum).  A single vortex at $\nu = \frac{1}{2}$ will have charge $-1/2$. Thus the electron bound with the single vortex will have charge-$ + 1/2$. We thus obtain a dipole with two  $2\pi$ vortices at either end, one with electric charge $+1/2$ and the other with electric charge $-1/2$  (see Fig. \ref{newdip}).

This dipole picture is very close to the ones developed before. It however makes clear how particle-hole symmetry operates and captures the essential features of Son's proposed description. To see this cleanly consider the limit in which the two ends of the dipole are separated by a distance much larger than the ``size" of each vortex. Then the self and mutual statistics of the two ends of the  dipole are well defined. One end carries a $2\pi$ vortex and an associated electric charge 1/2, and hence is a semion.  The other end of the dipole   is an anti-semion as it carries a $2\pi$ vortex but now with opposite electric charge $-1/2$.  They clearly are also mutual semions (see Fig. \ref{dippiphs}), {\em i.e}, when one of these goes around the other there is a phase of $\pi$.  In the absence of this mutual statistics, the dipole - as a bound state of a semion and an anti-semion - will be a boson. However the mutual statistics converts this bound state into a fermion,  exactly consistent with direct expectations since we are binding two vortices to the electron. 

Let us now turn to the action of $C$. Note first that as the electric charge is odd under $C$, while the vorticity is even, the effect of $C$ is to reverse the direction of the relative coordinate ({\em i.e}, the dipole moment). 
This should be contrasted with the standard picture where the dipole moment is reversed under $C$ but the $4\pi$ vortex is unaffected so that the particle-hole transformed object is not simply related to the original one.  

We can now understand the Kramers doublet structure (under $C$) directly from this new picture of the particle-hole symmetric composite fermion.    Let us fix one end of the dipole to be at the origin, and understand the dynamics of the relative coordinate.   Due to the phase $\pi$ when the relative coordinate rotates by $2\pi$, the orbital angular momentum is quantized to be a half-integer.   If we restrict to the low energy doublet with orbital angular momentum $\pm \frac{1}{2}$, the orientation of the relative coordinate become the $x$ and $y$ components of a spin operator $\vec S$ that acts on this doublet.  The $z$-component is then the angular momentum $\pm \frac{1}{2}$ of the two states in the doublet. As both this angular momentum and the relative coordinate are odd under $C$, we have  $C \vec S C^{-1} =  - \vec S$. It follows immediately that this dipole is a Kramers doublet.  

Finally it is easy to argue that these are Dirac fermions. Though at zero momentum the two states in the doublet are degenerate, at any non-zero momentum, there will be a dipole moment as explained above. 
In the new theory the dipole moment is precisely the $x, y$ components of the ``spin" of the Kramers doublet - so the locking of the dipole moment to the direction perpendicular to the momentum is precisely the spin-momentum locking  of a Dirac fermion. In particular if the momentum is rotated by $2\pi$ the dipole moment rotates by $2\pi$ and the wave function has a phase of $\pi$.

These arguments are spelt out in detail in Appendix \ref{drcdip}.  There we solve a simple  problem  of two quantum particles of opposite charge moving in a uniform magnetic field.  The two particles are taken to be mutual semions, {i.e}, when one goes around the other there is a phase of $\pi$. Further we impose an anti unitary  $C$ symmetry that interchanges the coordinates of the two particles. 
The solution shows the emergence of both the Kramers structure as well as the spin-momentum locking of the dipolar bound state of these two particles. 

If we form a Fermi surface of these composite fermions, the low energy state at any momentum point $\vec K$ will have a unique direction of ``spin" polarization perpendicular to $\vec K$. It's Kramers partner is the state at $- \vec K$ which has exactly the opposite ``spin" polarization. When the composite fermion goes around it's Fermi surface the rotation of the momentum by $2\pi$ thus forces a Berry phase of $\pi$.

We can see that this `new' dipole is the  natural fate of the `old' dipolar picture when $\nu = 1/2$ and particle hole symmetry is taken into account.    
 
Thus we now have a very simple physical picture of the structure of the particle-hole symmetric composite fermion.  
This physical picture also establishes a continuity between  the theory of the particle-hole symmetric composite fermi liquid with the earlier descriptions.   

We turn next to a different understanding of the particle-hole symmetric half-filled Landau level which yields powerful insights.

 \section{The half-filled Landau level as a topological insulator surface state}
 \label{hlltisrf}
It is important to emphasize that the $C$ symmetry at $\nu = \frac{1}{2}$ is not an exact ultra-violet (UV) symmetry of the theory. Further  it does not act locally in the microscopic Hilbert space. It is an emergent non-local symmetry of just the lowest Landau level at half-filling with the restriction to a two-body interaction (or more generally to $2n$-body terms).  As a matter of principle an exact projection from higher Landau levels will also have three-body terms, etc which will break the $C$ symmetry.  A useful approximation, in the limit of weak Landau level mixing, is to ask about the ground state in the lowest Landau level with exact $C$ symmetry, and then understand the $C$-breaking effects as a perturbation. 

Can we find a UV completion of the half-filled  Landau level that retains $C$ as an exact microscopic local symmetry? We turn next to this question.

Consider fermions in $3d$ with a symmetry group $U(1) \times C$. For now we define $C$ acting on these fermions to be an anti unitary operator which is such that the generator of the $U(1)$ symmetry is odd under $C$. 
As an example consider a lattice tight binding Hamiltonian 
\begin{eqnarray}
H_{3d} & =  & \sum_{ij}\sum_s t_{ij} c^\dagger_{is} c_{js}  + h.c  \nonumber \\
& & + \Delta_{ij} \left(c^\dagger_{i\uparrow} c^\dagger_{j\downarrow} + c^\dagger_{i\downarrow} c^\dagger_{j\uparrow} \right) + h.c \nonumber
\end{eqnarray}
Here $i, j$ are sites of a $3d$ lattice, $s = \uparrow, \downarrow$ is the electron spin. The triplet Cooper pairing term breaks charge conservation, and $SU(2)$ spin rotations but leaves a $U(1)$ subgroup of rotations generated by $S^z$ invariant. So long as the hopping and pairing parameters are real the Hamiltonian is also invariant under an anti unitary time reversal operation which we denote $C$ that acts locally and takes $c_{is} \rightarrow i\left(\sigma_y \right)_{ss'} c_{is'}$.  

Consider gapped free fermion Hamiltonians with this symmetry\footnote{This symmetry class is denoted $A III$ in the topological insulator literature.} .  The progress on topological insulators/superconductor shows that in $3d$ such systems are classified\cite{schnyder08,kitaev08} by the group $Z$ corresponding to an integer topological invariant which we label $n$. Correspondingly at the two dimensional interface with the vacuum there is a gapless surface state with $n$ Dirac cones with the Lagrangian: 
\begin{equation}
\label{surfdirac}
{\cal L} = \sum_{\alpha = 1}^n \bar{\psi}_\alpha \left(-i\slashed{\partial }\right) \psi_\alpha
\end{equation}
with the following symmetry action
\begin{eqnarray}
U(\lambda) \psi_\alpha U^{-1}(\lambda) & = & e^{i\lambda} \psi_\alpha \\
C\psi_\alpha C^{-1} & = & i\sigma_y \psi_\alpha^\dagger
\end{eqnarray}
The fermions $\psi_\alpha$ are each $2$-component and the corresponding $\gamma$ matrices are $\gamma_0 = \sigma_y, \gamma_1 = \sigma_z, \gamma_2 = \sigma_x$.  The fermion density $\psi_\alpha^\dagger \psi_\alpha$ is odd under $C$. Thus the symmetry action on the surface is $U(1) \times C$ as required. Further the oddness under $C$ implies that we cannot add a chemical potential term so that the Dirac fermions are necessarily at neutrality.

Recent work\cite{3dfSPT2,maxvortex} shows that with interactions this $Z$ classification is reduced to $Z_8$ (so that only $n = 0, 1, ...., 7$ are distinct phases)\footnote{There is an additional Symmetry Protected Topological phase which cannot be described within free fermion theory so that the full classification\cite{3dfSPT2} is $Z_8 \times Z_2$.}.  We will henceforth focus on the $n = 1$ state which is stable to interactions. 

We will take the liberty of calling the generator of the global $U(1)$ symmetry as `charge' irrespective of its microscopic origins in an electron model. This charge is odd under the anti unitary $C$ operation.  We will further take the liberty of occasionally referring to $C$ as ``time reversal".   When the results are applied to  the half-filled Landau level discussed in the previous section the $C$ operation will be interpreted physically precisely as the anti-unitary particle-hole symmetry transformation (hence the same symbol as in the previous section). In that context $C$ should of course not be confused with physical time reversal which is not a 
symmetry of the half-filled Landau level. 

Consider coupling the surface theory, at $n = 1$, to external static ``electromagnetic" fields that couple to the $U(1)$ charge and current densities. As the charge  is odd under $C$ the current is even. Then electric fields are $C$-odd while magnetic fields are $C$-even. We can thus perturb the surface theory by introducing an external magnetic field while preserving the $U(1) \times C$ symmetry.   We will work in a limit in which we assume that the continuum approximation (Eqn. \ref{surfdirac}) is legitimate. The resulting Lagrangian takes the form
\begin{equation}
{\cal L} = \bar{\psi} \left(-i\slashed{\partial } + \slashed{A}\right) \psi + ....
\end{equation}
with $\vec \nabla \times \vec A = B \hat{z}$ (taking the surface to lie in the $xy$ plane). 
The $.....$ represent four fermion and other interaction terms consistent with symmetry. In the absence of these interactions the spectrum has the famous Dirac Landau levels with energy $E_m = \pm \sqrt{2mB}$.  For non-zero $m$ each level comes with a partner of opposite energy.  Most importantly there is a zero energy Landau level that has no partner. Now the $C$ symmetry implies that this zeroth Landau level must be half-filled. 

At low energies it is appropriate to project to the zeroth Landau level. We thus end up with a half-filled Landau level. As usual in the non-interacting limit this is highly degenerate and we must include interactions to resolve this degeneracy.   Thus the surface of this $3+1$-d topological insulator maps exactly to the classic problem of the half-filled Landau level. Note however that the $U(1) \times C$ symmetry of the full TI maps precisely to the expected $U(1) \times C$ symmetry of the half-filled Landau level.

Thus we have obtained a UV completion that retains $U(1) \times C$ as an exact microscopic local symmetry.  The price we pay is that it is the boundary of a TI that lives in one higher dimension.  {\em Further our ability to obtain it this way implies that there is no strictly $2d$ UV completion of the half-filled Landau level that has $U(1) \times C$ as an exact local symmetry.} 

It follows that to understand the half-filled Landau level we must study strongly correlated surface states of the $n = 1$ $3+1$-dimensional topological insulator with $U(1) \times C$ symmetry.  
 
 \section{Correlated surface states of 3d topological insulators}
 \label{cfsti}
Let us consider quite generally the surface of a three dimensional topological insulator. To keep continuity with the previous section we will phrase the discussion in terms of the $n = 1$ $3+1$-D topological insulator with $U(1) \times C$ symmetry. We also initially specialize to $B = 0$. Later we will turn on a non-zero $B$. The simplest surface state - and the only one realized within band theory - is the free Dirac cone described by the Lagrangian in Eqn. \ref{surfdirac} with $n = 1$.  With interactions though other states are possible\cite{avts12,fSTO1,fSTO2,fSTO3,fSTO4,neuperttisrf2015,mrosscdl15}. 

The surface may spontaneously break the defining symmetries. For instance if $C$ is broken, then a Dirac mass is allowed. This leads to a quantized Hall conductance which is shifted from integer by a $1/2$.  Thus if we consider a domain wall between the two possible orientations of the $C$-breaking order parameter, it will support a chiral current carrying edge mode.  Crucial to the discussion that follows will be a different surface state that preserves $C$ but spontaneously breaks the global $U(1)$ symmetry - a surface `superconductor'. Finally a gapped surface that preserves the full $U(1) \times C$ symmetry is also possible. This price to pay is that such a surface state has what is known as `intrinsic topological order' with gapped `anyon' excitations carrying fractional charge. For the $n = 1$ topological  insulator of interest such a state was described in Refs. \onlinecite{fidkowski3d,3dfSPT2,maxvortex}, and shown to be non-abelian. We will return to this state later but first we discuss the surface superconductor in greater detail. 
 
We will restrict attention to gapped superconducting ground states. As is usual in any superconductor the excitations are gapped fermionic Bogoliubov  quasiparticles, and vortices which quantize external magnetic flux in units of $\frac{mh}{2e} \equiv m\pi$. In addition in the absence of long range Coulomb interactions, there is a gapless zero sound (Goldstone) mode which leads to a logarithmic interaction between the vortices. We will initially ignore this zero sound mode - later we will be able to reinstate it in a straightforward manner. 
 
 This superconducting state preserves the $C$ symmetry, and we can ask about the $C$ transformation properties of the various excitations.  As the $U(1)$ charge is odd under $C$,  the phase of the Cooper pair is even under $C$. It follows that the vorticity is even under $C$. The structure of the vortices has many similarities to those in the familiar Fu-Kane superconductor\cite{fuknsc} obtained at the surface of the usual topological insulator.  In particular $m\pi$ vortices $v_m$ with $m$ odd trap Majorana zero modes. As we are imagining turning off the coupling to the zero sound mode (for instance by weakly coupling the $U(1)$ currents to a gauge field), the vortices will have finite energy and we can discuss their statistics. Due to the Majorana zero modes $v_m$ with $m$ odd will be non-abelian. 
 
 What about $v_m$ with $m$ even? Below we will argue that there are two vortices at $m = 2$ denoted $v_{2\pm}$ one of which is a semion and the other is an antisemion. These two differ by binding a neutralized Bogoliuobov quasiparticle. They also go into each other under the $C$ operation. Most crucially from these we can build a $m = 4$ vortex that goes to itself under the $C$ operation by binding together $v_{2+}$ and $v_{2-}$. Remarkably this bound state which we dub $v_4$ is a fermion that is a `Kramers doublet' under the anti-unitary $C$ operation: 
 \begin{equation}
 C^2 v_4 C^{-2} = -v_4
 \end{equation}
 We can also construct other $m = 4$ vortices by binding the neutralized Bogoliuibov quasiparticle to $v_4$ (they can be thought of as $v_{2+}^2 \sim v_{2-}^2)$.  Finally the strength-$8$ vortex is a boson that transforms trivially under $C$.

\begin{figure}
\begin{center}
\includegraphics[width=2.5in]{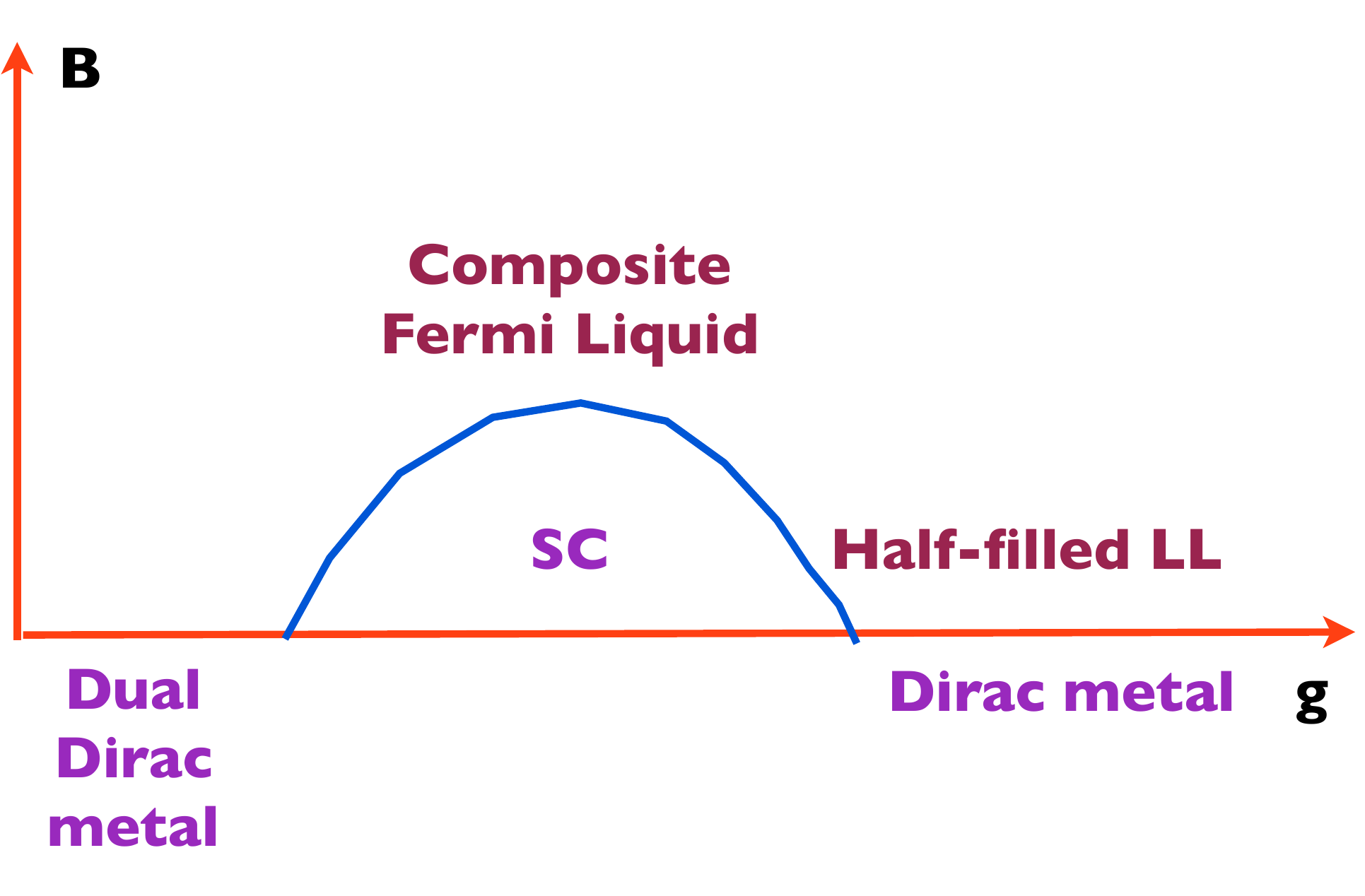}
\end{center}

\caption{Schematic phase diagram for the surface of the correlated $n = 1$ topological insulator with $U(1) \times C$ symmetry. $g$ is a parameter that controls the relative strength of Cooper pairing versus the phase stiffness in the superconductor. At $B = 0$ with increasing $g$ pairing is lost leading to the Dirac metal obtained within band theory. At small $g$ superconductivity is destroyed through phase fluctuations leading to the `dual' Dirac metal. Many other phases are also of course possible. Particularly interesting is a symmetry preserving surface topological order.  At non-zero $B$, the composite fermi liquid emerges as one of the possible  phases.}
\label{tisrfpdccl}

\end{figure}

 We will justify these results in the following section. But for now we pause to describe our strategy for understanding correlated surface states (including when a non-zero $B$-field is turned on).  We start from the surface superconductor and ask how the broken $U(1)$ symmetry may be restored.  One option is that the superconducting order is destroyed by losing the pairing gap. At $B = 0$  this leads to the free Dirac cone, and at $B > 0$ to the half-filled Landau level whose fate will be decided by interactions. Alternately we may destroy the superconducting order through phase fluctuations, {\em i.e} by proliferating vortices.  To obtain a symmetry preserving state we must proliferate vortices that are either fermions or trivial bosons. The former leads to gapless surface states. In particular when $B > 0$ it leads to a quantum vortex liquid of $v_4$ vortices. The resulting state is remarkably similar to the composite Fermi liquid expected in the half-filled Landau level with the additional virtue that it is manifestly $C$ symmetric.  We depict in Fig. \ref{tisrfpdccl} a schematic phase diagram of the surface of this TI illustrating some of the various possibilities. 
 
 \section{Topological insulators and time reversal symmetric $U(1)$ quantum spin liquids}
 \label{titsymmu1}
 How should we understand the claims made about the structure of the even strength vortices in the surface superconductor?  One approach is to work directly with the surface theory and examine the structure of the vortices in greater detail. Within the Bogoliubov-deGennes mean field theory, strength-$m\pi$ vortices will have $m$ Majorana zero modes. Knowing the action of $C$-symmetry, we can then study the fate of these zero modes in the presence of interactions to deduce the properties of the vortices. Here however we will describe a different and more insightful approach which enables us to deduce the properties of the even strength vortices. 
 
\begin{figure}
\begin{center}
\includegraphics[width=2.5in]{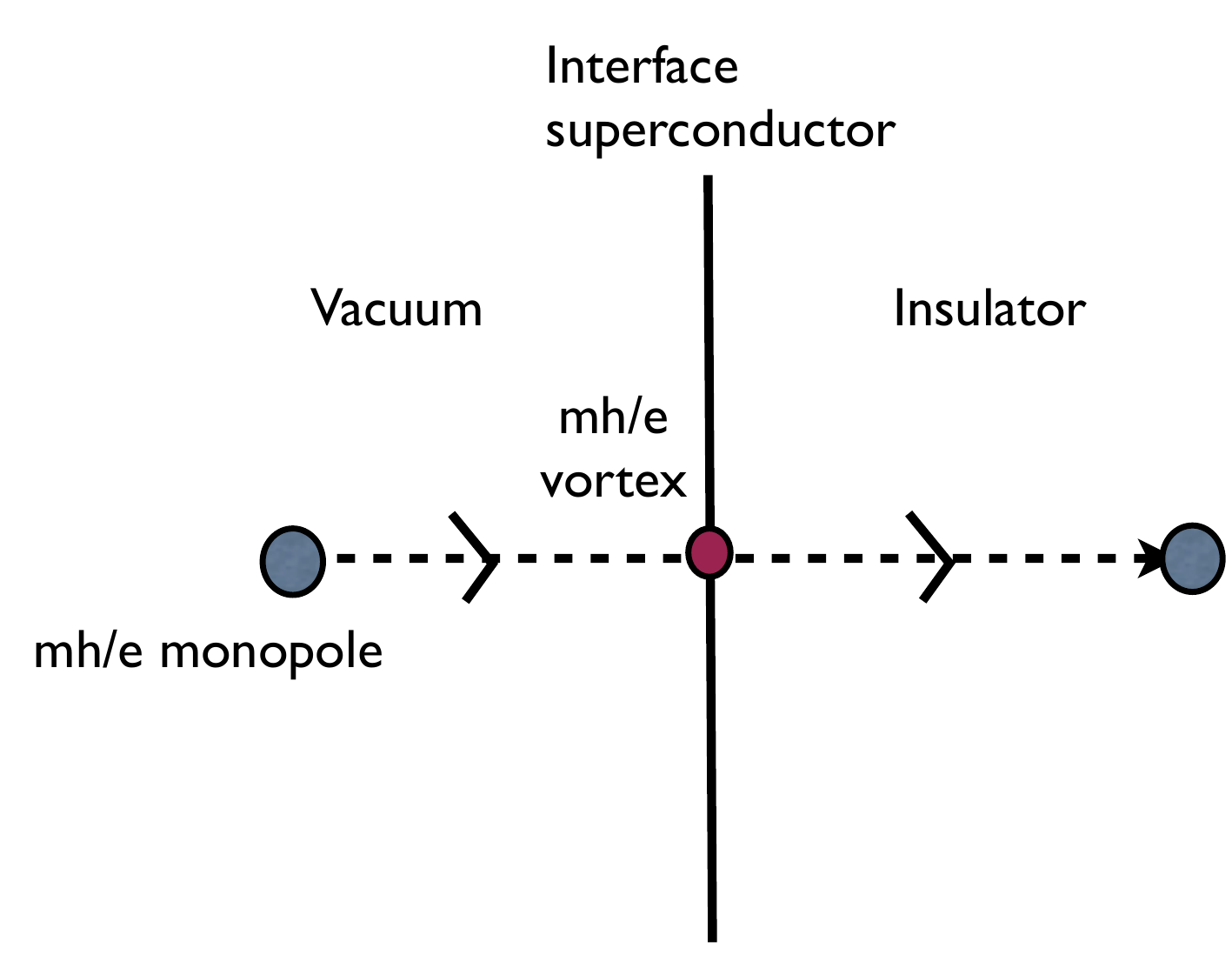}
\end{center}

\caption{When a strength $\frac{mh}{e}$ monopole from the outside vacuum tunnels into the bulk insulator through a surface superconductor, it leaves behind a strength ${mh}{e}$  vortex at the surface. Understanding the properties of the monopole in the bulk constrains the properties of the corresponding surface vortex.}
\label{mnpltnn}

\end{figure}

 First let us ask how we might create such vortices in the first place.  As usual a strength $\frac{mh}{e} \equiv m\pi$ vortex with even $m$ may be created in a superconductor by threading in external magnetic flux of $\frac{mh}{e}$ through a point. At the surface of the three dimensional bulk this process of flux insertion has a very nice and useful interpretation. We can think of it as throwing a magnetic monopole from the outside vacuum into the sample of the topological insulator as depicted in Fig. \ref{mnpltnn}. Recall that by Dirac quantization the magnetic monopole has strength $\frac{mh}{e}$ with $m$ even. When such a monopole passes through the superconducting surface to enter the bulk it leaves behind precisely a $m\pi$ vortex with $m$ even. 
 
 Thus the properties of the surface vortices can be inferred from the properties of the  bulk magnetic monopoles\cite{statwitt,fSTO1,3dfSPT2} or vice versa\cite{hmodl,3dfSPT2,maxvortex}.  In the outside vacuum the monopole is a trivial boson. If inside the topological insulator sample the monopole has some non-trivial properties then the vortex left behind at the surface through a monopole tunneling event will also inherit the same non-trivial properties. We emphasize that at this stage the bulk monopole is a `probe' of the system and should not be viewed as a dynamical excitation. 
 
To discuss these monopoles somewhat precisely let us imagine that we couple the microscopic fermions that form the bulk insulator to a dynamical compact $U(1)$ gauge field in its deconfined phase. The microscopic fermions become the elementary electric charges. We are interested in the fate of the magnetic monopoles. 

Consider a bit more carefully the interpretation of the theory obtained by gauging an insulator formed out of the fermions. Note that the fermions themselves are not local degrees of freedom in such a theory. To create a fermion we also need to create the electric field lines that emanate from it and go out to infinity. More formal a single electric charge creation operator is by itself not gauge invariant. Gauge invariant local operators are bosonic combinations made out of bilinear (or other even numbers of the fermions). Thus it follows that after gauging the theory should be regarded as living in the Hilbert space of a spin or boson system. 

In the last three decades it has been appreciated\cite{wenbook} that systems of interacting quantum spins/bosons can settle into quantum spin liquid phases characterized by emergent gauge fields and associated matter fields with fractional quantum numbers. In three dimensional systems it has long been recognized\cite{wenbook,bosfrc3d,wen03,3ddmr,hfb04,lesikts05,kdybk} that quantum spin liquid phases exist where there is an emergent `photon' excitation which is gapless with a linear dispersion. In addition there will be particle-like excitations that couple to the photon as  electric/magnetic charges. These states of matter are called $U(1)$ quantum spin liquids to emphasize that their low energy physics is described by an emergent deconfined $U(1)$ gauge theory. 

The phase obtained by gauging an insulator of fermions should thus be viewed as a particular kind of $U(1)$ quantum spin liquid. Since the fermions are gapped the electric charges in this spin liquid are gapped. Further in the gauged theory the global $C$ symmetry is still present. Thus we have an example of a $U(1)$ quantum spin liquid `enriched' by the presence of a global anti-unitary $C$ symmetry.  As noted above we could equally well simply call $C$ as ``time-reversal". Thus the discussion that follows can be usefully understood as being about time reversal symmetric $U(1)$ quantum spin liquids in three space dimensions.  

Consider obtaining an effective low energy Lagrangian for the  photon by integrating out the matter fields in such a  spin liquid.  Quite generally the $C$ symmetry implies that this  takes the form
\begin{equation}
{\cal L}_{eff} = {\cal L}_{Max} + {\cal L}_\theta
\end{equation}
The first term is the usual Maxwell term and the second is the `theta' term:
\begin{equation}
{\cal L}_\theta = \frac{\theta}{4\pi^2} \v{E}\cdot\v{B}
\end{equation}
where $\v{E}$ and $\v{B}$ are the electric and magnetic fields respectively. 

As is well known (see in the TI context the reviews in Refs. \onlinecite{HsnMr,QiScz}  the $C$ symmetry  restricts the allowed values to $\theta$ to an integer multiple of  $\pi$.  When the electrically charged fermions form the $n = 1$ topological band discussed above it is easy to argue that $\theta = \pi$. This follows for instance from the shift by $1/2$ of the surface `integer' quantum Hall effect obtained when time reversal is broken\cite{HsnMr,QiScz}. Let us now understand the implications of this for the monopole structure which is our primary interest in this section. 

\begin{figure}
\begin{center}
\includegraphics[width=2.5in]{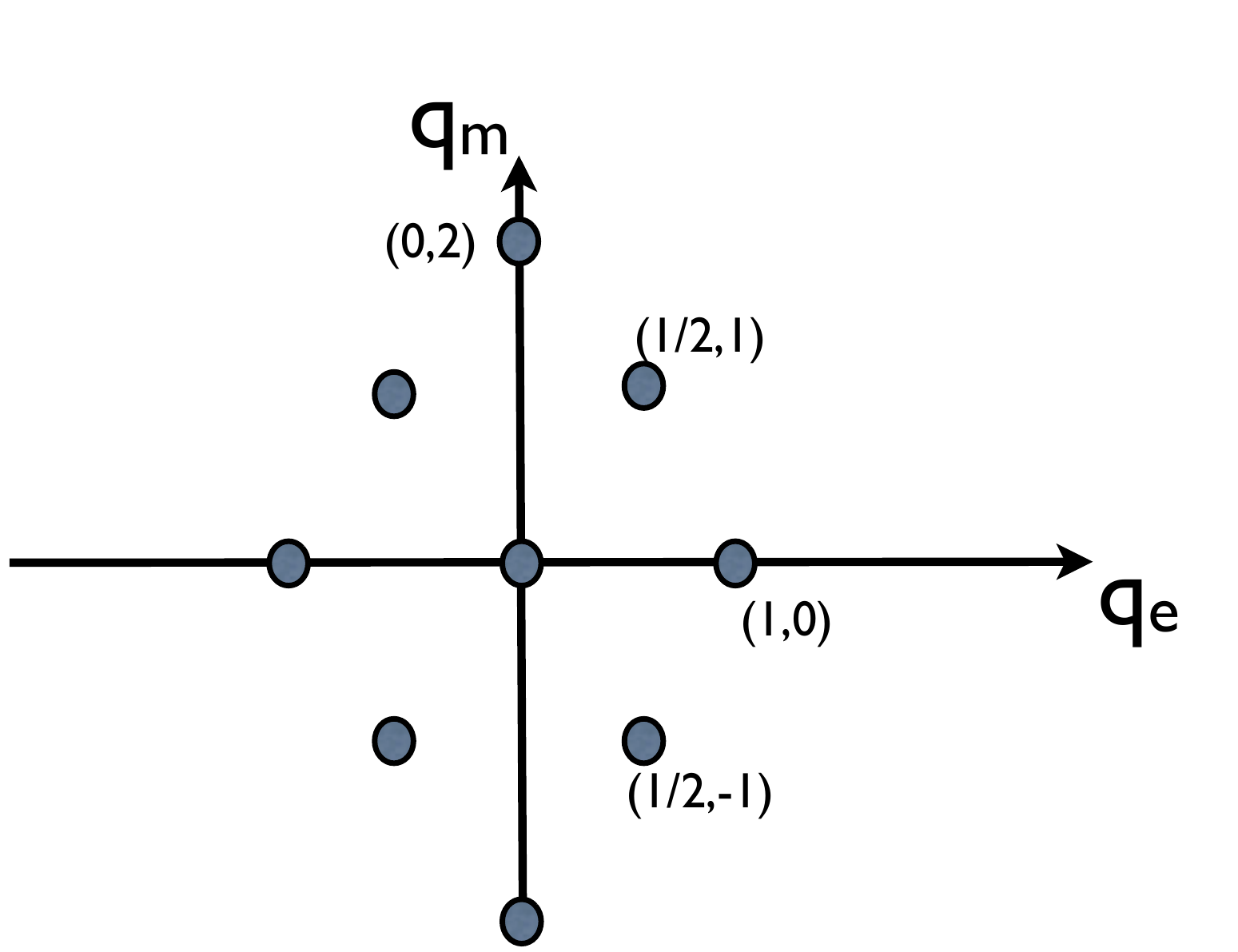}
\end{center}

\caption{ Charge-monopole lattice at $\theta = \pi$. }
\label{cmlat2}

\end{figure}

In the presence of a $\theta$ term, monopoles with magnetic charge $q_m = 1$ (we define the magnetic flux to be $\frac{hq_m}{e} \equiv 2\pi m$) carry electric charge $\pm \frac{1}{2}$ through the famous Witten effect\cite{wittendyon}. 
It will also be necessary to consider higher strength monopoles. To that end consider generally the lattice of allowed  electric and magnetic charges in this $U(1)$ gauge theory.  We will call this the charge-monopole lattice. It takes the form shown in Fig. \ref{cmlat2}.  

Let us denote by $(q_e, q_m)$ the electric and magnetic charges of the various particles, and by $d_{(q_e, q_m)}$ the corresponding destruction operator. 
We have chosen units in which the elementary `pure' electric charge is $(1,0)$. This particle is a fermion. The elementary strength-$1$ monopoles are then $(\pm \frac{1}{2}, 1)$ particles with bose statistics. The $(1,2)$ dyons are clearly also bosons as they are obtained by binding two $(\frac{1}{2}, 1)$ dyons. However the electrically neutral $(0,2)$ particle is a fermion. It can be obtained by removing an elementary fermionic electric charge ($(1,0)$ particle) from the $(1,2)$ dyon. 

It is actually extremely useful to construct the $(0,2)$ dyon differently as the bound state of the $(\frac{1}{2}, 1)$ and $(-\frac{1}{2}, 1)$ dyons. These two dyons see each other as mutual monopoles. Suppose one of these dyons, say the $(-\frac{1}{2}, 1)$ is sitting at some point we define to be the origin. When the other dyon, the $(-\frac{1}{2}, 1)$, traces out a  a loop  it picks up a phase equal to half the solid angle subtended by the loop at the origin. Binding two such bosonic particles produces a fermion\cite{asgdyon}. 

Consider now the action of the $C$ symmetry. As the electric charge is $C$-odd the magnetic charge must be $C$-even. It follows that $C$ interchanges the $(\pm \frac{1}{2}, 1)$ dyons. Thus in their bound state the relative coordinate is odd under $C$. Now  the angular momentum stored in the electromagnetic field of this bound state is readily calculated to be $\frac{1}{2}$, {\em i.e} the bound state behaves as a ``spin-1/2" particle as expected from the Fermi statistics. Further in this spin-$1/2$ Hilbert space the unit vector along the relative coordinate becomes precisely the spin operator.  As this is $C$-odd the two degenerate states of this spin-$1/2$ form a Kramers doublet. More details are in Appendix \ref{blkbd}. Specifically
\begin{equation}
C^2 d_{(0,2)} C^{-2} = - d_{(0,2)}
\end{equation}
The structure of other dyons can be easily obtained from these few basic ones. We will not need them here. 

In passing we note the strong similarity between this discussion and that in Section \ref{phcflphys} (and Appendix \ref{drcdip}) where we argued for the Kramers structure of the dipole in Fig. \ref{newdip}. This similarity is not coincidental: as we will see there is a deep connection between the bulk $(0,2)$ dyon and the surface composite fermion. 

Armed with this understanding  let us discuss the vortices in the surface superconductor.  Consider first the $2\pi$ vortices. These may be understood as points of penetration at the surface of $2\pi$ magnetic flux lines that extend into the bulk. Now the $\theta = \pi$ term in the bulk implies that when two closed bulk  $2\pi$ flux lines link there is a phase of $(-1)$. 
This linking phase ensures that when a single $2\pi$ flux line is cut open to produce a strength-$1$ monopole it costs infinite energy unless it binds to $\pm \frac{1}{2}$ electric charge.  The binding to the electric charge removes the linking phase ambiguity of an open flux tube and enables the resulting $(\pm \frac{1}{2}, 1)$ dyon to have finite energy, exactly consistent with  the Witten effect. 

We can now infer the statistics of the $2\pi$ vortices at the surface. When one  such vortex is taken around one another, the change in the flux line configuration can be deformed to an extra pair of linked flux lines in the bulk. Thus when a $2\pi$ vortex is taken around another there is a phase of $\pi$.  Note that corresponding to the two bulk dyons $(\pm \frac{1}{2}, 1)$ we will have two surface $2\pi$ vortices $v_{2\pm}$. The $\pi$ phase is picked up when any of these $2\pi$ vortices goes around the other.  This implies that these $v_{2+}$ and $v_{2-}$ are mutual semions, and that their self-statistics is either semion or anti-semion. Further since in the bulk the two dyons are interchanged by $C$, the same will be true for $v_{2 \pm}$ at the surface.  It follows that one of them ($v_{2+}$ )must be a semion and the other $v_{2-}$  an anti-semion.

Now let's discuss $4\pi$ vortices.  When a strength $q_m = 2 $ monopole tunnels through the surface from the vacuum into the bulk it leaves behind a $4\pi$ vortex.  We have already seen that in the bulk the $(0,2)$ monopole is a fermion that is Kramers doublet under $C$. It follows that at the surface there is a $4\pi$ vortex - which we dub $v_4$ - which is a Kramers doublet (under $C$) fermion. 

Thus thinking about the bulk gives us a simple understanding of the claims made in the previous section about the surface vortices. The $v_4$ vortex will play a crucial role in the discussion that follows. 

 Further understanding of the surface superconductor is provided by the considerations of the next section.  
 
 \section{Bulk duality of the gauged topological insulator}
 \label{bdgti}
 We now argue that the $U(1)$ quantum spin liquid obtaining by gauging the $n = 1$ $U(1) \times C$ topological insulator has a remarkable dual description\cite{tsymmu1,dualdrcmaxav}.    
 First of all we know that the charge-monopole lattice has the structure shown in Fig. \ref{cmlat2}. 
The most fundamental particles in this lattice are the  $( \frac{1}{2}, \pm 1)$ dyons. All other particles can be obtained as composites of these. Let us first discuss their statistics.  As they are  interchanged under $C$, they are required to have the same statistics, {\em i.e} they are both bosons or both fermions.  Further we already observed that the  $((\frac{1}{2}, 1)$ and the $(\frac{1}{2}, -1)$ dyon are relative monopoles, {\em i.e} each one sees the other the way an electric  charge sees a monopole.  If these dyons were both fermions  we would have a realization of an  ``all-fermion" $U(1)$ gauge theory in a strictly $3+1$ dimensional system. However it has been argued in Ref. \onlinecite{3dfSPT} (see also Ref. \onlinecite{kmcgsw14}) that such a state cannot exist.  Therefore we conclude that both these dyons must be bosons.

We have already also argued that the bound state - the $(0,2)$ particle - of these two dyons is a Kramers doublet fermion. Now consider the pure electric charge - the $(1,0)$ particle -  obtained by binding 
 $(\frac{1}{2}, 1)$ and $(\frac{1}{2}, -1)$. 
These are also relative monopoles and hence their bound state is a fermion. Now $C$ does not interchange these two dyons and hence the argument above for the Kramers structure of the $(0,2)$ particle does not apply.  

Earlier we obtained this phase by starting with fermionic electric charges forming the $n = 1$ topological band and gauging it. The present discussion shows that fermi statistics of the electric charge is {\em necessary} to realize this charge-monopole lattice. 

We thus see that the structure of both the elementary electric charge and the elementary magnetic charge are uniquely determined for this charge-monopole lattice. In addition the statistics and symmetry properties of the elementary dyons is also fixed. Thus there is a  unique possibility for this charge-monopole lattice.

Consider now this charge-monopole lattice from the point of view of the $(0,2)$ Kramers doublet fermion.  This is the elementary pure magnetic charge in this spin liquid.  Dirac quantization demands that the dual electric charge be quantized in units of half-integers. In this charge-monopole lattice the elementary electric charge with $q_e = \frac{1}{2}$ also necessarily has magnetic charge $q_m = 1$ which is exactly half that of the elementary pure magnetic charge. Thus as seen by the $(0,2)$ Kramers fermion there is also a dual Witten effect. This implies that this $(0,2)$ fermion itself is in a  topological insulator phase. As the magnetic charge is $C$ even this topological insulator is the same as the conventional topological insulator in spin-orbit coupled electronic insulators in three dimensions\footnote{These arguments show the uniqueness of the bulk excitation structure. There can still potentially be different phases distinguished by their possible surface states\cite{tsymmu1}. These are obtained by combining the spin liquid with a SPT phase of spins protected by time reversal alone. For the particular spin liquid discussed here, this subtlety is resolved in Metlitski (unpublished) and there is no extra such SPT  phase added in the duality}. 
. 

Thus the same phase admits two equivalent but dual points of view. We can obtain it either by taking the $n = 1$ topological insulator of fermions with $U_e(1) \times C$ symmetry and gauging the global $U_e(1)$ or by taking the standard topological insulator of Kramers fermions with $U_m(1) \rtimes C$ symmetry\footnote{This simply means that the generator of the $U_m(1)$ is even under $C$. As $C$ is anti-unitary this implies that $U_m(1)$ rotation and $C$ do not commute.} and gauging this $U_m(1)$.   For clarity in this section  we use the subscripts $e$ or $m$ for $U(1)$ to distinguish between the `electric' and `magnetic' $U(1)$ rotations. 

\section{Duality of surface states}
\label{dualsrf}
It is interesting to translate this bulk duality into a dual perspective of the surface states.  The simplest case is the superconducting surface.    
We recall that the surface avatar of the $(0,2)$ monopole is the $v_4$ vortex.  We thus seek a dual description of the superconducting state in terms of the physics of the $v_4$ vortex. 

Let us first quickly review  pertinent aspects of the standard charge-vortex duality of two dimensional systems\cite{chandandual,mpafdhl89}. The simplest example is for bosonic superfluids. Then the superfluid phase may be fruitfully viewed as a Mott insulator of vortices in the phase of the boson. The zero sound mode of the superfluid can conveniently be represented as a gapless photon in $2+1$ dimensions, and the vortices couple to this photon as `electric charges'. . This leads to a dual Landau-Ginzburg theory of the superfluid in terms of vortex fields coupled minimally to a fluctuating non-compact $U(1)$ gauge field. The magnetic flux of this gauge field corresponds physically to the physical boson number density.

It has also been known for some time now\cite{z2long} how to extend this dual vortex formulation to an ordinary gapped $s$-wave superconductor of {\em fermions} in two dimensions.  To describe the Bogoliubov quasiparticle it is convenient to formally strip them of their electric charge and define neutralized fermionic particles (``spinons") which see the elementary $\frac{h}{2e}$ vortices as $\pi$-flux. The vortices are in addition coupled minimally, exactly as in a bosonic superfluid, to a fluctuating  non-compact $U(1)$ gauge field. This dual description of an ordinary superconductor is conceptually powerful, and enables passage from the superconductor to various fractionalized Mott insulators in two dimensions. 

Returning now to the superconductor obtained at the surface of the $n = 1$ TI with $U(1) \times C$ symmetry, from the point of view of the $v_4$ vortex the surface is gapped. Further the vortex number conservation is the surface manifestation of the magnetic $U_m(1)$ gauge structure present in the bulk spin liquid. The preservation of the vortex number conservation means that the surface preserves the dual $U_m(1) \rtimes C$ symmetry. 
Thus from the point of view of $v_4$ what we have been calling the surface superconductor is really a symmetry preserving surface topological order of the bulk topological insulator formed by the $(0,2)$ fermions. 

It is possible to check this explicitly. Following the logic described in the previous section we can fully determine the braiding/fusion rules, and the symmetry assignment  for the quasiparticles of the surface superconductor. These turn out to be identical to that of a specific surface topological order (known as T-Pfaffian\cite{fSTO3}) obtained earlier through bulk Walker-Wang constructions for the spin-orbit coupled topological insulator with the $v_4$ identified with the dual `electron' (and thus a vorticity $4\pi$ identified with dual `electron' charge $1$). 

We are now ready to describe the full dual Landau-Ginzbuirg theory of the surface superconductor by reinstating the zero sound mode. As usual this zero sound mode is described as a gapless photon in $2+1$ dimensions. The vortices will then couple minimally to this photon.  Thus a dual Landau-Ginzburg description of the surface superconductor is simply obtained: Take the T-Pfaffian topological order and couple all the charged particles to a fluctuating non-compact $U(1)$ gauge field. $a_\mu$ (Recall that the charges of the T-Pfaffian are precisely the vortices of the surface superconductor).  

This dual formulation of the surface superconductor will be extremely useful as a framework in which to address non-superconducting states obtained through phase fluctuations. We turn to these next.

 \section{Vortex metal surface states}
 \label{vms}
 The surface superconducting order may be destroyed to restore $U(1) \times C$ symmetry by proliferating vortices. If we condense bosonic vortices, for instance the $8\pi$ vortex $v_4^2$, we will get a symmetry preserving gapped surface topological order.  Alternately we can kill the superconductivity by proliferating the fermionic $v_4$ vortex, {\em i.e} by making it gapless.   As the dual LGW theory of the surface superconductor is the gauged version of the T-Pfaffian topological order, we will get a gapless vortex liquid if we confine the non-trivial quasiparticles of the T-Pfaffian state through a phase transition to  a gapless symmetry  preserving state of the $v_4$ fermion. But this is precisely the famous single Dirac cone (tuned to neutrality) formed by $v_4$. We thus have a dual Dirac liquid surface state\cite{dualdrcwts2015,dualdrcmaxav} for the $n = 1$ $U(1) \times C$ topological insulator described by the Lagrangian
 \begin{equation}
 \label{ddl}
 {\cal L} = \bar{\psi}_v \left(-i\slashed{\partial} - \slashed{a}\right) \psi_v + \frac{1}{4\pi} \epsilon_{\mu\nu\lambda} A_\mu\partial_\nu a_\lambda
 \end{equation}
 Here $\psi_v$ is a fermion field representing the $v_4$ vortex. We have chosen units so that this couples to the non-compact  gauge field $a_\mu$ with gauge charge-$1$. With this choice the conserved 3-current of the  original global $U(1)$ symmetry is 
 \begin{equation}
 \label{dualjagain}
 j_\mu = \frac{1}{4\pi} \epsilon_{\mu\nu\lambda} \partial_\nu a_\lambda
 \end{equation}
 This is reflected in the last term of the Lagrangian which describes the coupling of this current to the external probe gauge field $A_\mu$.  Finally the original electron $\psi$ is obtained as $4\pi$ instanton in the gauge field $a_\mu$. 
 Importantly $\psi_v$ is {\em Kramers doublet} under the $C$ operation transforming as 
 \begin{equation}
 C\psi_v C^{-1} = i\sigma_y \psi_v
 \end{equation}

 This dual Dirac liquid describes a possible surface state if the surface superconductivity is destroyed by phase fluctuations at zero magnetic field $B$. What if the superconductivity is destroyed by turning on a non-zero $B$? Now we will  have a finite density of vortices. If we wish to preserve $C$ symmetry the simplest option is to induce a finite density of $v_4$ vortices and make them form a `metallic' state. This will lead to a non-zero chemical potential in the Lagrangian in Eqn. \ref{ddl}  for the dual Dirac liquid so that the dual Dirac cone is no longer tuned to be at the neutrality point. The density of these vortices is precisely 
 \begin{equation}
 \label{nvdens}
 n_v = \frac{B}{4\pi}
 \end{equation}
 as these are $4\pi$ vortices. 
 Further as these are fermions they will form a Fermi surface.  The Fermi momentum $K_F$ will be related to $n_v$ in the usual way
 \begin{equation}
 K_F = \sqrt{4\pi n_v}
 \end{equation}
 The fermions at this Fermi surface will of course continue to be coupled to the $U(1)$ gauge field $a_\mu$.

 \section{Back to Composite Fermi Liquids}
 \label{bcfl}
 Let us now return to the fate of the half-filled Landau level in the presence of particle-hole symmetry. Earlier we argued that we can UV complete this theory with the $U(1) \times C$ symmetry retained as an exact locally realized microscopic symmetry by obtaining it as the surface of the $n = 1$ TI with $U(1) \times C$ symmetry.  We now see that when $B \neq 0$ at this surface, as required to produce the half-filled Landau level, a possible gapless state that preserves the $U(1) \times C$ symmetry is the dual Dirac liquid at non-zero chemical potential. 
 
 This theory bears some remarkable similarities to the usual composite Fermi liquid description. We will therefore identify the field $\psi_v$ (or equivalenty  the $v_4$ vortex) with the composite fermion. First the density of $\psi_v$ as given by Eqn. \ref{nvdens} is precisely half the degeneracy of the lowest Landau level, {\em i.e} it matches exactly the density of electrons in the half-filled Landau level. Just as in the usual composite fermi liquid, $\psi_v$ forms a Fermi surface which is then coupled to a non-compact $U(1)$ gauge field. $\psi_v$ itself is formally electrically neutral (it is a vortex) but the gauge flux couples to the external vector potential. 
 
The main difference is that particle-hole symmetry is explicitly present in this version of the composite Fermi liquid. Further $\psi_v$ is a Kramers doublet under $C$, and its Fermi surface encloses a Dirac cone. This is manifested in a $\pi$ phase when a $\psi_v$ particle at the  Fermi surface circles around it. 

This is precisely the description of the particle-hole symmetric composite fermion liquid proposed by Son in Ref. \onlinecite{sonphcfl} which we described in Sec. \ref{sonprop}. We have thus provided an understanding of Son's proposal through the linkage with the surface of a three dimensional electronic topological insulator.  

It is worth emphasizing a few points. The vortex metal/composite fermi liquid surface state has been shown to emerge as a legitimate surface state of the $n = 1$ topological insulator with $U(1) \times C$ symmetry in a non-zero $B$-field. This same surface  also provides a realization of a half-filled Landau level with $U(1) \times C$ symmetry. Thus the vortex metal/composite fermi liquid state is a legitimate state for a half-filled Landau level with particle-hole symmetry. Whether this state is really the fate of the half-filled Landau level or not depends on microscopic details which we have not attempted to address. 

A different question altogether is whether the dual Dirac liquid at zero field describes the same phase as the standard single Dirac cone.  We have also not attempted to answer this question here.

\begin{figure}
\begin{center}
\includegraphics[width=3.3in]{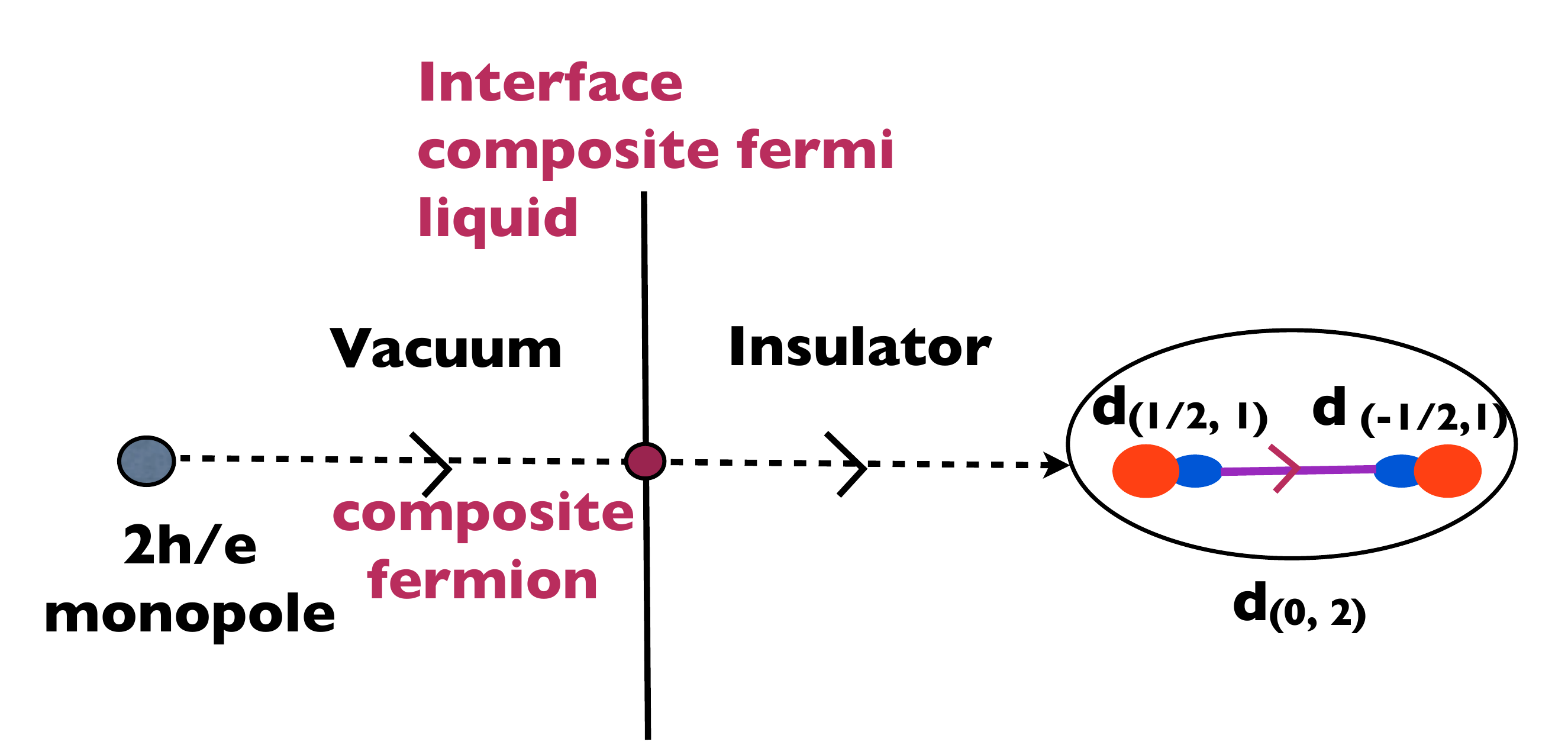}
\end{center}

\caption{Bulk-boundary correspondence for the composite fermi liquid. The composite fermion is the surface avatar of the electrically neutral strength-$2$ bulk monopole which itself is a bound state of the two $(\pm \frac{1}{2}, 1)$ dyons. This strength-$2$ monopole is a fermion, and is Kramers-doublet under $C$. At the surface the two dyons that make up this monopole correspond to the two ends of the dipole of Fig. \ref{newdip}.}
\label{blkbdrcfl}

\end{figure}

Finally we discuss the physical  picture of the composite fermion from the point of view of the three dimensional topological insulator when the half-filled Landau level is obtained as its boundary. Note that the composite fermion is the surface avatar of the strength-$2$ electrically neutral monopole in the bulk (see Fig. \ref{blkbdrcfl})..  We earlier obtained the properties of this strength-$2$ monopole by obtaining it as the bound state of the $(\pm \frac{1}{2}, 1)$ dyons. These two dyons correspond precisely, at the surface,  to the two oppositely charged $2\pi$ vortices at the two ends of the composite fermion. 

 \section{Particle-hole symmetric Pfaffian state}
 \label{phpf}
The  composite fermi liquid state is well known to act as a `parent' normal state out of which the non-abelian Moore-Read (Pfafian) state arises through pairing\cite{readgrn}.  It is also well known\cite{ssletalapf,levinapf} that the Pfaffian state breaks particle-hole symmetry. A particle-hole conjugate state - known as the anti-Pfaffian- has been described as an alternate candidate for the observed plateau at $\nu = \frac{5}{2}$. From the particle-hole symmetric composite fermi liquid, it is natural then to consider angular momentum $l = 0$ pairing which preserves  the particle-hole symmetry. This leads to a gapped topologically ordered state - which we may call the $C$-Pfaffian - which is yet another alternate possible non-abelian quantum Hall state at the same filling. 

It is interesting to view this as a correlated surface state of the related three dimensional topological insulator with $U(1) \times C$ symmetry. As it preserves the $U(1) \times C$ symmetry, this is a symmetry preserving 
surface topological order. Precisely such surface topologically ordered states were described in Refs. \onlinecite{fidkowski3d,3dfSPT2,maxvortex}. The $C$-Pfaffian state obtained by $l = 0$ pairing of the composite fermions of the particle-hole symmetric composite fermi liquid is essentially identical to the states described in these references.  

We briefly describe the particle content of the $C$-Pfaffian state.  In the absence of the gauge field $a_\mu$ this is simply the famous Fu-Kane superconductor obtained at the surface of spin-orbit coupled $3+1$-d topological insulators.  In particular the fundamental $\pi$ vortex (and all odd multiples)  traps a Majorana zero mode.  The presence of the gauge field means that the vortices are screened and will have finite energy cost. The $\pi$ vortex (and its odd multiples) will clearly have non-abelian statistics. Through Eqn. \ref{dualj} we see that the $\pi$ -vortex will have physical electric charge $\frac{e}{4}$.  An argument identical to the one in 
Sec. \ref{titsymmu1} shows that there are two  $2\pi$ vortices, carrying charge $\pm \frac{e}{2}$, one of which is  a semion and the other an antisemion.  These are also mutual semions.  Their bound state is a $4\pi$ vortex which is an electrically neutral fermion, and is a Kramers doublet under $C$. This is exactly the Bogoliubov quasiparticle obtained after the pair condensation of composite fermions. As usual this Bogoliubov quadiparticle has $\pi$ mutual statistics with the $\pi$ vortex but is local with respect to the $2\pi$ vortices. 

A full description of the braiding and fusion rules and other topological data is readily obtained for the $C$-Pfaffian state. We however here focus on showing the connection with the physical picture described in the previous sections of  the modified dipolar picture of the composite fermion (see Fig. \ref{newdip}). We already emphasized that the neutral fermion of the $C$-Pfaffian state was Kramers under $C$, and should be understood as the relic of the composite fermion. We also see that it can be understood as the bound state of the charge $\frac{e}{2}$ semion and the charge $ -\frac{e}{2}$ antisemion.  But this is precisely the dipolar picture advocated in the previous section. In particular the two ends of the dipole have been liberated as deconfined quasiparticles by the passage to the paired $C$-Pfaffian state. This lends further support for this dipolar picture. 

It is also enlightening to relate the structure of the $C$-Pfafian state to the properties of the bulk $3+1$-D topological insulator with $U(1) \times C$ symmetry.  Then the neutral fermion of the $C$-Pfaffian is precisely the surface avatar of the strength-$2$ electrically neutral magnetic monopole. The charge $\pm \frac{e}{2}$ anyons (either semion or antisemion) are the surface avatars of the $(\pm \frac{1}{2}, 1)$ dyons. This ties in beautifully with the pictures described in previous sections.

\section{Revisiting the phenomenology of composite fermi liquids}
\label{cflphen}
With the understanding of the half-filled Landau level described above  it is interesting to revisit the phenomenology of composite fermi liquids (with or without particle-hole symmetry). By and large these are unchanged from the original HLR theory. We also describe some new experimental predictions (that do not actually rely crucially on particle-hole symmetry).

In practice even if the projection  to the lowest Landau level and the restriction to two-body interactions is a good approximation, there will inevitably be disorder potentials that will break particle-hole symmetry. Further the edge potential also breaks particle-hole symmetry so that physical quantities sensitive to edge physics will not be particle-hole symmetric.   

Nevertheless in an ideal sample, if Landau level mixing can be neglected, we expect the formulation described here will apply. In that case  how can the $\pi$-Berry phase associated with the Fermi surface of the composite fermions be measured? We show in Appendix \ref{sdhpi} that this $\pi$ Berry phase is implied already by a slight re-interpretation of the standard phenomenology away from $\nu = \frac{1}{2}$.

\begin{enumerate}
\item
{\em Electromagnetic response} 

The electromagnetic response functions of the $C$-symmetric composite Fermi liquid were discussed in Ref. \onlinecite{sonphcfl}. They resemble but are not identical to those proposed by the standard HLR theory. To discuss dc transport at low-$T$ it is necessary to include the effects of disorder. A random potential will, as in the standard HLR theory, lead to a random magnetic field seen by the composite fermions.  For simplicity we assume that the probability distribution of the random potential is particle-hole symmetric. Then the mean effective field seen by the composite fermions is zero. Consider the electrical conductivity tensor.  When we access the half-filled Landau level as a TI surface state we have to include a contribution to the Hall conductivity of $\frac{e^2}{2h}$ from the filled states below the chemical potential.  To understand this precisely, note that when the lowest Landau level is obtained in the usual way in two dimensional systems, the empty Landau level has Hall conductivity $0$ and the filled one has Hall conductivity $\frac{e^2}{h}$. However when this Landau level  is obtained as the surface state of a $3d$ TI, the empty and full levels are related by $C$-symmetry: they hence have opposite Hall conductivities $\pm \frac{e^2}{2h}$ respectively. Thus the surface Dirac composite fermion theory must be supplemented by the background term (Eqn. \ref{lbg}) described in Sec. \ref{sonprop} when describing the usual Landau level.

Thus the full physical conductivity tensor $\sigma_{ij}$ takes the form 
\begin{equation} 
\sigma_{ij} = \frac{e^2}{2h} \epsilon_{ij} + \sigma^*_{ij}
\end{equation}
Here $\epsilon_{ij}$ is antisymmetric and $\epsilon_{xy} = 1$. $\sigma^*_{ij}$ is the conductivity tensor calculated within the low energy effective field theory given by Eqn. \ref{ccfl}. A physical description of this conductivity is easily obtained. First there is no off-diagonal term in $\sigma^*$ as the $\psi_v$ move in zero effective magnetic field. Second the $\psi_v$ are $4\pi$ vortices in the electron phase. Thus by the usual rules of  charge-vortex duality the electrical conductivity of the vortices is proportional to the inverse of their 
resistivity obtained within the standard RPA.  More precisely we have 
\begin{equation}
\label{nsigma}
\sigma^*_{ij} = \delta_{ij} \frac{e^4}{(4 \pi \hbar )^2 \sigma_v}
\end{equation}
Here $\sigma_v$ is the RPA expression for the conductivity of the $\psi_v$ composite fermions (we have reinstated factors of $e$ and $\hbar$).  It follows that the measured physical longitudinal conductivity is just 
\begin{equation}
\label{sigmaxx}
\sigma_{xx} = \frac{e^4}{(4\pi \hbar)^2 \sigma_v}
\end{equation}
As a function of wavenumber $q$, the composite fermion conductivity $\sigma_v$ takes the well-known form 
\begin{eqnarray}
\sigma_v & = & \frac{e^2K_F l}{4\pi \hbar}, ~~ q \ll \frac{2}{l} \\
& = & \frac{e^2 K_F}{2\pi \hbar q}, ~~ q \gg \frac{2}{l}
\end{eqnarray}
where $l$ is the impurity induced mean free path for the composite fermions. Combining with Eqn. \ref{sigmaxx} the physical longitudinal conductivity takes exactly the form obtained by HLR in their original theory, and used to confront a number of experiments\cite{dassarmabook}. 

Note that in the usual HLR theory there is a composition rule for the resistivity (rather than the conductivity) tensor: 
\begin{equation}
\label{orho}
\rho_{ij}^{HLR} = \frac{2h}{e^2} \epsilon_{ij} + \rho^*_{ij} 
\end{equation}
where $\rho^*$ is the resistivity tensor of the composite fermions. In practice we are in the limit $\rho_{xx} \ll \rho_{xy}$, and further $\rho_{xy}$ is approximately $\frac{2h}{e^2}$ even in the standard theory. 
Thus in HLR theory the longitudinal conductivity 
\begin{equation}
\sigma_{xx}^{HLR} \approx \frac{e^4 \rho^*_{xx}}{4h^2 }
\end{equation} 
which is essentially the same as Eqn. \ref{sigmaxx} (after identifying $\rho^*_{xx}  \approx \frac{1}{ \sigma_v}$).

\item
{\em Thermal transport and Wiedemann-Franz violation}

A striking feature of conventional Fermi liquid metals is the Wiedemann-Franz relationship between the residual electrical and thermal conductivities.  Within Boltzmann transport theory, in the limit $T \rightarrow 0$, the longitudinal thermal conductivity  $\kappa_{xx}$ is related to the electrical conductivity through 
\begin{equation}
\label{convwf}
\kappa_{xx} = L_0 T \sigma_{xx}
\end{equation}
where $L_0 = \frac{\pi^2k_B^2}{3e^2}$ is the free electron Lorenz number.    

We now argue that the composite Fermi liquid will not satisfy the conventional Wiedemann-Franz law but will instead satisfy a modified one. Though the composite fermions contribute to electrical transport as vortices, they 
are directly responsible for heat transport.  Thus the measured residual $\kappa_{xx}$  will satisfy Wiedemann-Franz with the $\sigma_v$, {\em i.e} the composite fermion conductivity.  But this is inversely related to the measured electrical conductivity. Thus we have the relation 
\begin{equation}
\label{modwfsigma}
\kappa_{xx} = \frac{L_0 T e^4}{4 h^2\sigma_{xx}}
\end{equation}
Conceptually similar violations have been discussed previously\cite{vmetal} in other vortex metals.  Equivalently we observe that the longitudinal resistivity is, to a good approximation which ignores corrections of order  $\left(\frac{\rho_{xx}}{\rho_{xy}}\right)^2$, given by 
\begin{equation}
\rho_{xx} = \frac{(4h)^2 \sigma_{xx}}{e^4}
\end{equation}
so that the modified Wiedemann-Franz law may be be written
\begin{equation}
\label{modwfrho}
\kappa_{xx} \rho_{xx} = L_0 T
\end{equation}
If instead we use the standard HLR theory we will obtain Eqn. \ref{modwfrho} as an essentially exact relation (so long as we can ignore off-diagonal terms in $\rho^*$) and Eqn. \ref{modwfsigma} will hold approximately upto ignoring corrections of order $\left(\frac{\rho_{xx}}{\rho_{xy}}\right)^2$.

For a   conventional metal in {\em zero} magnetic field, the modified Wiedemann-Franz law(Eqn. \ref{modwfrho}) is equivalent to the usual one as $\sigma_{xx} = \frac{1}{\rho_{xx}}$. However in a non-zero magnetic field, Eqns. \ref{convwf} 
and \ref{modwfrho} are no longer equivalent. 

For a conventional metal in non-zero magnetic field Eqn. \ref{convwf} is the appropriate result (more generally the thermal conductivity tensor is equal to $L_0 T$ times the electrical conductivity tensor)\cite{Abrikosovbook}.  However for the composite Fermi liquid Eqn. \ref{modwfsigma} (or Eqn. \ref{modwfrho}) holds. 

It is interesting to quantify the violation of the conventional Wiedemann-Franz law by defining a Lorenz number $L_{CF}$ for the composite Fermi liquid through
\begin{equation}
L_{CF} = \frac{\kappa_{xx}}{T\sigma_{xx}}
\end{equation}
We have 
\begin{equation}
\frac{L_{CF}}{L_0} = \left(\frac{\rho_{xy}}{\rho_{xx}}\right)^2
\end{equation}
Since the measured $\rho_{xx} \ll \rho_{xy}$ we have a giant enhancement - possibly of order $10^3$ - of the Lorenz number compared to free electrons.

This modified Wiedemann-Franz law can possibly be tested in experiments. We emphasize that this result does not rely on particle-hole symmetry and is indeed obtained in  the standard HLR theory as well.  Similar violations are expected at $\nu = \frac{1}{4}$ and other composite fermi liquid metals. We are not aware of any thermal conductivity measurements in the $\nu = \frac{1}{2}$ state.  Of course it will be necessary to subtract off the thermal conductivity of the substrate. This can perhaps be done by comparing with the thermal conductivity at a neighboring quantum Hall plateau\footnote{The off-diagonal thermal conductivity $\kappa_{xy}$ 
will however satisfy the conventional Wiedemann-Franz with the electrical $\sigma_{xy}$ so that $\kappa_{xy} = L_0 T \sigma_{xy}$. This means that $\kappa_{xx} \gg \kappa_{xy}$ so that the longitudinal thermal {\em resistivity} $\approx \frac{1}{\kappa_{xx}}= \frac{\rho_{xx}}{L_0 T}$.  This form of the Wiedemann-Franz law is also equivalent to Eqn. \ref{convwf} at zero field but becomes inequivelent in non-zero field. }.

\item
{\em Cyclotron orbits} 

If the Landau level filling is changed from $1/2$, particle-hole symmetry will be broken. Just like in the original HLR theory, the composite fermions will see an effective magnetic field that is much reduced from the externally applied one. They will then have cyclotron orbits with a radius much bigger than for electrons in the same external magnetic field. 

%Let us imagine we move away from half-filling by changing the electron density by $\delta \rho$ at a fixed external field. The electron filling is then changed by $\delta \nu = \frac{2\pi \delta \rho}{B}$.  As $B$ is fixed the density of composite fermions is fixed by Eqn. \ref{nvdens}. However these now see an effective internal magnetic field 
% \begin{equation}
 %\label{bstar}
 %B^*= - 4\pi \delta \rho~~= - 2B \delta \nu
% \end{equation}
 %The cyclotron radius $R^*_c$ for the composite fermions is thus given by 
 %\begin{eqnarray}
 %R^*_c & = & \frac{K_F}{|B^*|} \nonumber \\
 %& = & \frac{1}{2|\delta \nu| B}
 %\end{eqnarray}
% Note that $R^*_c$ decreases symmetrically as $\nu$ is changed in either direction from $1/2$ but the ratio $R^*_c |\delta \nu|$ does not change. 
 
Consider moving away from half-filling by changing the magnetic field by $\delta B$ while keeping the electron density fixed. The filling is changed to $\delta \nu = -\frac{\delta B}{2B}$.    The deviation from half-filling  changes $\langle j_0 \rangle$ through Eqn. \ref{jorho} to 
\begin{equation}
\langle j_0 \rangle = -\frac{\delta B}{4\pi}
\end{equation}
Through Eqn. \ref{dualj} this is related to the average internal magnetic magnetic field. Thus the composite fermions see an effective magnetic field 
\begin{equation}
B^* =  \delta B ~~=  - 2B\delta \nu
\end{equation}

To leading order in $\delta \nu$,  the cyclotron radius of the composite fermions is
 \begin{equation}
 \label{rcstar1}
 R^*_c = \frac{K_F}{|B^*|} 
 \end{equation}
Thus we have 
\begin{equation}
R^*_c |B^*| = \frac{1}{l_B}
\end{equation}
where $l_B = \frac{1}{\sqrt{B}}$ is the magnetic length, which is the same result as in the standard HLR theory. 

Recently\cite{kamburov14} through a geometric resonance experiment, $R_c^*|B^*|$ was inferred as a function of a $\delta \nu$. The results were interpreted as indicating that 
$K_Fl_B$ decreased on deviating from $\nu = \frac{1}{2}$ in either direction. This has been addressed theoretically in Refs. \onlinecite{maissamph,brmtkjain15}. In the particle-hole symmetric theory, when the external magnetic field is  changed at fixed density, 
the density of composite fermions changes by $\delta n_v  = \frac{\delta B}{4\pi}$, and correspondingly the Fermi momentum changes by $\delta K_F = \frac{\delta B}{2\sqrt{B}}$.  From Eqn. \ref{rcstar1} this gives one source of $\delta \nu$ dependence which however leads to a steady decrease of $R_c^*|B^*|$ with increasing $\delta \nu$. However we caution that when $B^* \neq 0$ the composite fermion momenta are smeared on the scale of $\frac{1}{R_c^*} \sim \sqrt{B} |\delta \nu|$ which is the same order as $\delta K_F$. Thus the theory of the $\delta \nu$ dependence in the experiment likely requires more complicated analysis which we leave for the future. 

\item
{\em $2K_F$ density oscillations}

It is interesting to ask about the singularities in the $2K_F$ response of physical quantities in the particle-hole symmetric theory. Note that the physical charge density is not simply the composite fermion density (unlike in HLR). Since the physical density is given by  Eqn. \ref{dualj}, we see that the density correlator is determined by the correlator of the transverse gauge field. For simplicity let us specialize to zero frequency. Then, 
\begin{equation}
\langle |j_0(\vec q, \omega = 0 )|^2 \rangle  = q^2 \langle |a_t(\vec q, \omega = 0)|^2 \rangle
\end{equation}
where $a_t$ is the transverse component of the vector potential $\vec a$. For $q \approx 2K_f$ this means that the universal structure of the density correlator is the same as in that of the transverse gauge field. 
In the effective Lagrangian the gauge field couples to the fermions through the term
\begin{equation}
\bar{\psi}_v (\vec k + \vec q) a^i_{-q} \gamma^i \psi_v(\vec k) 
\end{equation}
For $\vec q \approx 2K_f \hat{x}$, the important coupling is between composite fermions in a patch of the Fermi surface near $+ K_F \hat{x}$ and those in an antipodal patch near $- K_F \hat{x}$.  As the ``spin" of the composite fermion is polarized perpendicular to the Fermi momentum the wavefunctions at the two antipodal Fermi points are orthogonal to each other.  This means that $a_x$ (which couples to $\sigma_y$) will not scatter a fermion from the right patch to the left one. However $a_y$ couples to $\sigma_x$ and  will be able to scatter composite fermions between these two patches. Thus the effective quadratic action for $a_y$ near wave vectors $\vec q \approx 2K_F \hat{x}$ will be determined by the correlations of $\psi_{vR}^\dagger \sigma_x \psi_{vL}$ (where $R,L$ refer to the right and left patches respectively). This will have the same structure of the $2K_F$ singularity as in a usual Fermi surface coupled to a gauge field \cite{mrossde,altshuleretal}.  In the presence of the long range Coulomb interaction these are essentially unmodified from the Fermi liquid form (upto log corrections) corresponding to a square root cusp as a function of $|q - 2K_F|$ that modifies a smooth non-universal contribution. It is easy to then see that the density correlations will have this same universal structure of $2K_F$ singularities (as in the standard HLR theory).

\item
{\em Disorder with statistical particle-hole symmetry: Localization} 

If the disorder is particle-hole symmetric we can ask about possible localization effects on the composite fermions. Ignoring the gauge field maps the problem to that of the surface Dirac cone of spin-orbit coupled $3d$ topological insulators in the presence of a random effective magnetic field $B_{eff}$ with statistical time reversal invariance. There will be some regions in space in which the magnetic field $B_{eft}$ is positive and some in which it is negative. Inside either of these regions if the magnitude of $B_{eff}$ is large there will be a gap and a $C$-broken gapped surface will be induced.  However along the domain walls between these regions there will be gapless $1d$ edge modes. In the strong disorder limit we will form a random network of these domain walls. We expect this to be at the critical point of the integer quantum Hall plateau transition.  (Similar arguments have been made in Ref. \onlinecite{fkwti} to discuss disorder effects on the surface of weak topological insulators, topological crystalline insulators, and related systems). This conclusion is presumably not affected by the gauge field. 

Thus statistically particle-hole symmetric disorder will not localize the composite fermions but rather drives the composite fermi liquid to the critical point of the integer quantum Hall plateau transition.

\end{enumerate}
  
  \section{Discussion}
  \label{disc}

We have elaborated in this paper the connections between three seemingly disparate research topics in quantum many body physics. Here we briefly comment on some extensions and open questions. 

For the half-filled Landau level, we presented various physical ways of understanding Son's proposed particle-hole symmetric theory.  This understanding will hopefully guide future efforts to derive the particle-hole symmetric composite fermi liquid theory by working purely within the lowest Landau level. For composite fermi liquids of bosons at $\nu = 1$ such a derivation was provided by Ref. \onlinecite{read98} building on the formulation of Ref. \onlinecite{Pasquier1998}.  For fermions at $\nu = \frac{1}{2}$ lowest Landau level approaches have been developed (see e.g., Ref. \onlinecite{gmrsrmp03}) but particle-hole symmetry has not been incorporated.

A different recent development\cite{kachru15} which  we did not describe here is the application of mirror symmetry of supersymmetric quantum field theories in $2+1$-d to the half-filled Landau level. Ref. \onlinecite{kachru15} started with  a  supersymmetric massless theory which is free in the infra-red, and which  is known to be dual to an interacting  supersymmetric gauge theory.   Turning on a magnetic field that couples to the conserved global $U(1)$ currents on the IR-free side of the duality breaks supersymmetry, and the low energy theory is simply that of a half-filled Landau level but for two species of fermions which couple with opposite electric charges to the external magnetic field. On the other side of the duality the effective gauge theory reduces essentially to Son's proposed theory but with two  fermi surfaces corresponding to the two species of fermions.

 In the introduction we raised the question of physical realization of correlated surface states of three dimensional topological insulators/superconductors. We now see that this has a surprising and interesting answer: a physical realization is the  half-filled Landau level of a two dimensional electron gas.  For topological insulators with $U(1) \times C$ symmetry with a $Z_8 \times Z_2$ classification, the $8$ distinct members of the $Z_8$ subgroup all have free fermion bulk realizations. The $n = 1$ member corresponds to the single half-filled Landau level. Higher values of $n$ are realized as multi-component quantum Hall systems where each component is at filling $\nu = \frac{1}{2}$. Such multicomponent systems have received a lot of attention over the years. We expect that the connection to topological insulator surface states will provide interesting insights just as it does for $n = 1$. 
 
The bulk duality of the gauged topological insulator has crucial implications for the classification and understanding of time reversal symmetric $U(1)$ spin liquids in $3+1$-dimensions. It shows that there is a unique such spin liquid where the low energy effective action for the emergent $U(1)$ gauge field has a $\theta$ angle of $\pi$. On the other hand when $\theta = 0$ Ref. \onlinecite{tsymmu1} showed that there were precisely $6$ distinct phases distinguished by the structure of bulk excitations leading to a total of $7$ distinct phases. Additional phases are obtained by combining these with SPT phases of the underlying spin system protected by time reversal.

We thank Patrick Lee for  strongly encouraging us to write this article, and for many discussions. We also thank M. Barkeshli, S. Das Sarma,  J. Eisenstein, Liang Fu, Matthew Fisher, N. Read, I. Sodemann, M. Metlitski,  M. Shayegan, D. Son, A. Vishwanath, Liujun Zou, and M. Zaletel for inspiring and informative discussions. This work was supported by NSF DMR-1305741.  This work was also partially supported by a Simons Investigator award from the
Simons Foundation to Senthil Todadri.  Since the submission of the initial version of this paper, two other papers (Refs. \onlinecite{geraedtsnum} and \onlinecite{msgmp15}) have appeared on the arxiv with further results on particle-hole symmetry  in the half-filled Landau level.

\appendix
\section{Dirac spectrum and Kramers structure for dipoles}
\label{drcdip}
Here we present some simple calculations which give much insight into how the primary physical features of the particle-hole symmetric composite fermion emerge out of the dipolar picture described in the main text. Consider two particles - one with charge $+q$ and the other with charge $-q$ - moving in a uniform magnetic field in two dimensions. The Hamiltonian is 
\begin{equation}
\label{dipH}
H = \frac{\vec \Pi_1^2}{2m} +  \frac{\vec \Pi_2^2}{2m} + V(\vec x_1 - \vec x_2)
\end{equation}
Here $\vec \Pi_1 = \vec p_1 - q \vec A(\vec x_1)$ , $\vec \Pi_2 = \vec p_2 + q \vec A(\vec x_2)$ are the kinematic momenta of the two particles. $\vec A$ is the vector potential corresponding to the uniform magnetic field $\vec B = B \hat{z}$, and $\vec x_{1,2}$ are the coordinates of the two particles. $V$ is an attractive interaction between the two particles. 

We will not repeat the solution of this problem here which has been studied a number of times over the decades (see Ref. \onlinecite{kallin84} and references therein). Our focus will be on a variant of this classic problem. First we  impose the condition that when one particle goes around the other , there is a phase of $\pi$, {\em i.e} the two particles are mutual semions. It is well known that the bound state then has Fermi statistics.  Here we are interested in a single composite particle formed by this binding (and the Fermi statistics is not directly relevant). Second we assume the existence of an anti-unitary symmetry operation $C$ that interchanges the two particles: 
\begin{equation}
C: \vec x_1 \leftrightarrows \vec x_2
\end{equation}

We will show that apart from having Fermi statistics a bound state of such a pair of oppositely charged particles in a magnetic field has all the essential properties of the particle-hole symmetric composite fermion discussed in the text, including the Dirac and Kramers structure. 

The mutual semion statistics imposes a restriction on the Hilbert space of wave functions $\psi(\vec R, \vec x)$ written in terms of center-of-mass $\vec R = \frac{\vec x_1 + \vec x_2}{2}$, and relative coordinates $\vec x = \vec x_1 - \vec x_2$.  Using polar coordinates $(r, \phi)$ for $\vec x$ we have
\begin{equation}
\psi(\vec R, r, \phi + 2\pi) = -\psi(\vec R, r, \phi)
\end{equation}
This will quantize the angular momentum conjugate to $\phi$ to be a half-integer. 

Following the standard solution of such a two-particle problem we define the two momenta
\begin{eqnarray}
\vec Q & = & \vec \Pi_1 + \vec \Pi_2 - q\vec x \times \vec B \\
\vec p & = & \frac{\vec \Pi_1 - \vec \Pi_2}{2}
\end{eqnarray}
It is straightforwardly checked that the pairs $(\vec R, \vec Q)$ and $(\vec x, \vec p)$ are canonically conjugate. Further we have $[Q_i, p_j] = [Q_i, x_j] = [Q_i, Q_j] = [p_i, p_j] = [R_i, p_j] = 0$.  It follows that $\vec Q$ commutes with the Hamiltonian (as is also obvious from the classical equations of motion).  The Hamiltonian in Eqn. \ref{dipH} may be rewritten
\begin{equation}
H = \frac{\left(\vec Q + q\vec x \times \vec B \right)^2}{4m} + \frac{\vec p^2}{m} + V(r)
\end{equation}

Note that under $C$, 
$\vec x$ and $\vec Q$ are odd while $\vec R$ and $\vec p$ are even, and $H$ is $C$-invariant.  We will take $V$ to only depend on the radial distance $r$ for simplicity.  
As $\vec Q$ commutes with the Hamiltonian (and its components commute with each other) we can fix it's value and consider just the relatiive motion.  
Expanding the first term out we get 
\begin{equation}
\frac{Q^2}{4m} + \frac{q\vec Q \times \vec x \cdot \vec B}{2m} + \frac{q^2B^2r^2}{4m}
\end{equation}
Due to the $\vec Q \times \vec x \cdot \vec B$ term the Hamiltonian is not, in general,  rotationally invariant, {\em i.e} $[\hat{l}, H] \neq 0$ where $\hat{l} = -i \frac{\partial}{\partial \phi}$ is the angular momentum operator. 
To illustrate the essential features of the bound state of the two particles we will begin by considering the special point $\vec Q = 0$ where there is rotational invariance:
\begin{equation}
H[\vec Q = 0] = \frac{1}{m} \left(-\frac{1}{r}\frac{ \partial}{\partial r} r\frac{ \partial}{\partial r} + \frac{l^2}{r^2} + \frac{q^2B^2r^2}{4} \right) + V(r)
\end{equation}
This Hamiltonian will have a bound state solution for the ground state. However since $l$ is quantized to be half-integer, there will be two degenerate ground states with $l = \pm \frac{1}{2}$ with some energy $E_0$.  Let us denote these two ground states $|\pm \frac{1}{2} \rangle$. Within this doublet ground state, the operator $l$ may be identified with $\frac{\sigma^z}{2}$ and $e^{i\phi}$ with $\frac{\sigma^x + i\sigma^y}{2}$ where $\vec \sigma$ are the standard Pauli matrices. Thus we introduce a ``spin-1/2" operator 
\begin{equation}
\vec S = (\cos \phi, \sin \phi, l)
\end{equation}

Consider now the action of $C$-symmetry.  As $\phi \rightarrow \phi + \pi$ under $C$, and $C$ is anti unitary, we have 
\begin{equation}
C \vec S C^{-1} = -\vec S
\end{equation}

It follows that the two-fold degenerate ground state is a Kramers doublet under $C$. Thus the degeneracy of this doublet is preserved under all perturbations that preserve $C$. 

Consider now the Hamiltonian at non-zero $\vec Q$: 
\begin{equation}
H[\vec Q] = \frac{Q^2}{4m} + \frac{q\vec Q \times \vec x \cdot \vec B}{2m}  + H[\vec Q = 0]
\end{equation}
A non-zero $\vec Q$ breaks the $C$ symmetry and hence the doublet will be split. However the spectrum will still be degenerate between $\vec Q$ and $-\vec Q$. To understand the splitting of the two-fold degeneracy we simply treat the non-zero $\vec Q$ terms in perturbation theory. Noting that in the doublet subspace $\vec x = \frac{r}{2}( \sigma_x, \sigma_y)$ we obtain - in first order perturbation theory and to linear order in $Q$ , an effective $2 \times 2$ Hamiltonian
\begin{equation}
H_{eff}[ \vec Q] = \frac{q\langle r \rangle \vec Q \times \vec S \cdot \vec B}{2m} + E_0
\end{equation}
This Hamiltonian (which depends on the ``center-of-mass" momentum $\vec Q$ and the spin $\vec S$ has the form of a Dirac Hamiltonian. In particular at any fixed $\vec Q \neq 0$, in the ground state the spin will be polarized to point along $\hat{Q} \times \hat{z}$.  This is the famous ``spin-momentum" locking expected of a two dimensional Dirac fermion. 

The states at $+\vec Q$ and $-\vec Q$ together form the two pairs of a Kramers doublet. At the $C$-invariant momentum $\vec Q = 0$ these pairs give a degenerate spectrum. 

%Consider now a many body state formed by a finite density of these bound states. As these are fermions we can fill a Fermi sea with states occupied upto some $|\vec Q| = K_F$. As the fermion   goes around a full circle in momentum space around the Fermi surface its momentum rotates by $2\pi$ and hence it's spin (which is rigidly locked perpendicular to the momentum) also by $2\pi$. Consequently the fermion wave function picks up a Berry phase of $\pi$. 

Note that the dipole moment ({\em, i.e}, the $x,y$ components of the ``spin") of the bound state is $\vec d = q\vec x = q \langle r \rangle \hat{Q} \times \hat{z}$. Thus the dipole moment/spin is indeed perpendicular to the momentum as expected. 

In this simple two-particle model there will in general be a non-zero term of order $Q^2$ (as readily established when the interaction $V = 0$) which will cause the dispersion to bend and eventually cross the energy at $Q = 0$  at some 
$|\vec Q| \sim \frac{1}{l_B}$ (where $l_B$ is the magnetic length).  This is presumably related to the non-anomalous implementation of $C$-symmetry in this model unlike the real half-filled Landau level. It is interesting to consider the limit $B \rightarrow \infty, m \rightarrow \infty$ such that $\frac{m}{\sqrt{B}}$ is finite. Then if we choose the zero of energy to coincide with the energy at $Q = 0$ we get a pure Dirac spectrum for all finite $Q$. 

\section{Structure of the $(0,2)$ bulk excitation}
\label{blkbd}
Here we briefly sketch the arguments\cite{3dfSPT,fSTO2} determining the structure of the bound state of the $(\frac{1}{2}, 1)$ and the $(-\frac{1}{2}, 1)$ dyons. This bound state is of course the $(0,2)$ particle.  First note that $(\frac{1}{2}, 1) = (-\frac{1}{2}, 1)+ (1,0)$.  Consider now a configuration where we have one dyon of either type.  The $(-\frac{1}{2}, 1)$ piece is common to both dyons and merely contributes an ordinary repulsive interaction. More profound is the effect of the remaining $(1,0)$ piece of the$ (\frac{1}{2}, 1)$ dyon on the other $(-\frac{1}{2}, 1)$ dyon.  Clearly the effect of this interaction is exactly the same as the interaction between $(1,0)$ and $(0,1)$ particles. 

It is well-known that when an electric charge $1$ moves in the potential of a  strength-$1$ monopole the angular momentum of the relative coordinate is quantized to be half-integer. 
The ground state has angular momentum $L = \frac{1}{2}$. In this ground state doublet the relative coordinate $\vec x_1 - \vec x_2 = \langle r \rangle \vec S$ where $\vec S$ is a spin-$1/2$ operator, {\em i.e} the orientation of the relative coordinate is precisely the spin operator of this ground state doublet. 

Now consider applying these well-known results to the system at hand.  The $C$-symmetry interchanges the two dyons so that $\vec x_1 \leftrightarrow  \vec x_2$.  Thus under $C$, in the ground state doublet $\vec S \rightarrow - \vec S$. As $C$ is anti unitary it follows that the two-fold degeneracy of the ground state is protected by Kramers theorem, and is insensitive to perturbations that preserve $C$ (even if spatial rotation is broken). 

Thus we conclude that the $(0,2)$ particle is a Kramers doublet. Further as a bound state of two bosonic particles which are mutual monopoles its statistics is fermionic. 

\section{Berry phase of $\pi$, Shubnikov-DeHaas oscillations, and the Jain sequence}
\label{sdhpi}

It is well known from studies of graphene and other Dirac materials that the $\pi$-Berry phase may be inferred experimentally by studying Shubnikov-DeHaas (SdH) or other quantum oscillations. For an ordinary  Fermi surface enclosing a Dirac node, the  SdH oscillations will show resistivity minima periodic at magnetic fields $B_n$ determined by 
\begin{equation}
\frac{1}{B_n} = \frac{n+ \frac{1}{2}}{F}
\end{equation}
where $n$ is an integer, and $F$ the frequency of oscillations is proportional to  the Fermi surface area. The shift by $\frac{1}{2}$ in the numerator is a consequence of the $\pi$-Berry phase and does not occur in a conventional Fermi surface. Thus a plot of the ``Landau index" $n$ versus $\frac{1}{B_n}$ will have an intercept $-\frac{1}{2}$. This is routinely used to detect the $\pi$-Berry phase.  Note that in a standard SdH experiment, the oscillations occur as a function of varying magnetic field at {\em fixed} density. 

Now let us turn to the composite Fermi liquid. We recall that in the formulation we are currently using the density of composite fermions and the effective magnetic field they see are given by 
\begin{eqnarray}
n_v & = & \frac{B}{4\pi} \nonumber \\
B^* & = & B - 4\pi \rho \nonumber
\end{eqnarray}
Thus to repeat the set-up of the standard SdH experiment we should keep the external $B$ fixed and move away from half-filling by tuning the density so that $n_v$ stays fixed but $B^*$ changes. 
Further  the longitudinal resistivity of the composite fermions is, through Eqn. \ref{sigmaxx}, proportional to the measured longitudinal conductivity $\sigma_{xx}$. However the measured resistivity $\rho_{xx}$ is also 
proportional to $\sigma_{xx}$ (as $\sigma_{xy} = \frac{e^2}{2h} \gg \sigma_{xx}$). It follows that the resistivity minima of the composite fermions track the minima of the measured resistivity. These (of course) occur at the filling of the Jain sequence
\begin{equation}
\label{jain}
\nu = \frac{n}{2n+1}
\end{equation}
At a fixed $B$, these correspond to values of the effective magnetic field 
\begin{equation}
\frac{1}{B_n^*} = \frac{2n +1}{B}
\end{equation}

It follows that a plot of $n$ versus $\frac{1}{B_n^*}$ will have an intercept $-\frac{1}{2}$ implying a $\pi$-Berry phase for the composite fermi surface. 

Resistivity minima at precisely the same fillings are of course a key feature of the standard HLR theory which does not associate any such Berry phase. How is this consistent? The point is that in HLR the composite fermion density is exactly equal to the electron density $\rho$ and not to $\frac{B}{4\pi}$. While these two are the same at $\nu = \frac{1}{2}$, they are different away from $\nu = \frac{1}{2}$. Then to think about the SdH oscillations we must keep $\rho$ fixed and tune $B$ to move away from half-filling. The Jain sequence (Eqn. \ref{jain}) then occurs at effective magnetic fields satisfying 
\begin{equation}
\frac{1}{B_n^*} = \frac{n}{2\pi \rho}
\end{equation}
Thus within the standard interpretation the plot of $n$ versus $\frac{1}{B_n}$ has zero intercept in agreement with the absence of a Berry phase.

\bibliography{cfltislrevbibl}

\end{document}